\documentclass[12pt]{article}
\usepackage{amsmath}
\usepackage{latexsym}
\usepackage{amssymb}
\usepackage{bm}

\newtheorem{theorem}    {Theorem}
\newtheorem{lemma}     {Lemma}
\newtheorem{remark}     {Remark}

\def\keywords{\vspace{-.3em}
    \if@twocolumn
      \small\it Keywords\/\bf---$\!$%
    \else
      \begin{center}\small\bf Keywords\end{center}\quotation\small
    \fi}
\def\endkeywords{\vspace{0.6em}\par\if@twocolumn\else\endquotation\fi
    \normalsize\rm}
\def\appendix{\par
    \setcounter{section}{0}\setcounter{subsection}{0}
    \def\thesection{\Alph{section}} \section*{Appendix}
}

\newcommand{\defeq}{\stackrel{\rm def}{=}}
\newcommand{\bR}{\mathbb{R}}

\newcommand{\bN}{\mathbb{N}}
\newcommand{\cH}{{\cal H}}
\newcommand{\cS}{{\cal S}}
\newcommand{\cC}{\mathfrak{C}}
\newcommand{\cP}{{\cal P}}
\newcommand{\cPn}{{\cal P}^{(n)}}
\newcommand{\cX}{{\cal X}}
\newcommand{\cXn}{{\cal X}^{(n)}}
\newcommand{\cXvec}{\vec{\cX}}
\newcommand{\cXnhat}{\hat{\cX}^{(n)}}
\newcommand{\cXvechat}{\vec{\hat{\cX}}}
\newcommand{\Tr}{{\rm Tr}\,}
\newcommand{\cHn}{{\cal H}^{(n)}}
\newcommand{\cHvec}{\vec{\cH}}
\newcommand{\rhon}{\rho^{(n)}}
\newcommand{\sigman}{\sigma^{(n)}}
\newcommand{\rhovec}{\vec{\bm{\rho}}}
\newcommand{\sigmavec}{\vec{\bm{\sigma}}}

\newcommand{\cCvec}{\vec{\cC}}
\newcommand{\cPvec}{\vec{\cP}}
\newcommand{\cSvec}{\vec{\cS}}
\newcommand{\Pn}{P^{(n)}}
\newcommand{\Pnhat}{\hat{P}^{(n)}}
\newcommand{\Pvec}{\vec{\bm{P}}}
\newcommand{\bW}{{\bf W}}
\newcommand{\bX}{{\bf X}}
\newcommand{\bY}{{\bf Y}}
\newcommand{\Wn}{W^{(n)}}
\newcommand{\Xn}{X^{(n)}}
\newcommand{\Yn}{Y^{(n)}}

\newcommand{\Kn}{K_n}

\newcommand{\cR}{{\cal R}}

\newcommand{\cFn}{{\cal F}_n}
\newcommand{\cFvec}{\vec{\cal F}}
\newcommand{\fn}{f_n}
\newcommand{\fvec}{\vec{\bm{f}}}

\newcommand{\Pe}{{\rm P_e}}

\makeatletter
\newcommand{\lleq}{\mathrel{\mathpalette\gl@align<}}
\newcommand{\ggeq}{\mathrel{\mathpalette\gl@align>}}
\newcommand{\gl@align}[2]{
\vbox{\baselineskip\z@skip\lineskip\z@
\ialign{$\m@th#1\hfil##\hfil$\crcr#2\crcr{}_{{}_{(=)}}\crcr}}}
\makeatother

\newcommand{\supD}{\overline{D}}
\newcommand{\infD}{\underline{D}}
\newcommand{\supJ}{\overline{J}}
\newcommand{\infJ}{\underline{J}}
\newcommand{\supI}{\overline{I}}
\newcommand{\infI}{\underline{I}}
\newcommand{\infH}{\underline{H}}
\newcommand{\supH}{\overline{H}}

\newcommand{\cHtensor}{{\cal H}^{\otimes n}}
\newcommand{\rhotensor}{\rho^{\otimes n}}
\newcommand{\sigmatensor}{\sigma^{\otimes n}}

\newcommand{\rhotvec}{\rhovec}
\newcommand{\sigmatvec}{\sigmavec}

\newcommand{\Ptn}{\Pn}
\newcommand{\cXtn}{{\cX}^n}
\newcommand{\Ptvec}{\Pvec}

\def\Label#1{\label{#1}\ [\ \verb+ #1 +\ ]\ }
\def\Label{\label}

\newcommand{\spec}[1]{\left\{#1\right\}}

\newcommand{\ch}{W}
\newcommand{\chn}{W^{(n)}}

\newcommand{\chvec}{\vec{\bm{W}}}
\newcommand{\charg}[1]{W_{#1}}
\newcommand{\chnarg}[1]{W^{(n)}_{#1}}
\newcommand{\chvecarg}[1]{\vec{\bm{W}}_{#1}}
\newcommand{\code}{\Phi}
\newcommand{\coden}{\Phi^{(n)}}
\newcommand{\codevarn}{\widetilde{\Phi}^{(n)}}
\newcommand{\codevec}{\vec{\bm{\Phi}}}
\newcommand{\dec}{Y}
\newcommand{\decn}{Y^{(n)}}
\newcommand{\decarg}[1]{Y_{#1}}
\newcommand{\decnarg}[1]{Y^{(n)}_{#1}}

\newcommand{\decvarn}{\widetilde{Y}^{(n)}}

\newcommand{\decvarnarg}[1]{\widetilde{Y}^{(n)}_{#1}}
\newcommand{\enc}{\varphi}
\newcommand{\encn}{\varphi^{(n)}}
\newcommand{\encarg}[1]{\varphi(#1)}
\newcommand{\encnarg}[1]{\varphi^{(n)}(#1)}

\newcommand{\xn}{x^n}
\newcommand{\xvec}{\vec{\bm{x}}}
\newcommand{\cost}{c}
\newcommand{\costn}{c^{(n)}}
\newcommand{\costvec}{\vec{\bm{c}}}

\newcommand{\Pnrc}{P^{(n)}_{\rm rc}}
\newcommand{\Enrc}{E^{(n)}_{\rm rc}}

\newcommand{\restrict}[1]{\!\!\restriction_{#1}}

\begin{document}
\title{General formulas for capacity of classical-quantum channels}

\author{
Masahito Hayashi
\thanks{
Laboratory for Mathematical Neuroscience,
Brain Science Institute, RIKEN,
2--1 Hirosawa, Wako, Saitama, 351--0198, Japan.
(e-mail: masahito@brain.riken.go.jp)
}
\and
Hiroshi Nagaoka
\thanks{
Graduate School of Information Systems,
University of Electro-Communications.
1-5-1, Chouhugaoka, Chouhu-shi, Tokyo, 182-8585, Japan.
(e-mail: nagaoka@is.uec.ac.jp)
}
}

\date{}

\maketitle

\begin{abstract}
The capacity of a classical-quantum channel (or in other words 
the classical capacity of a quantum channel) 
is considered in the 
most general setting, where 
no structural assumptions 
such as the stationary memoryless property are made on 
a channel.  
A capacity formula as well as 
a characterization of the strong converse property is 
given just in parallel with the corresponding classical 
results of Verd\'{u}-Han which are based on the so-called 
information-spectrum method.  The general results 
are applied to the stationary memoryless case with 
or without cost constraint on inputs, whereby a deep relation 
between the channel coding theory and the hypothesis testing 
for two quantum states is elucidated. 
\end{abstract}


\section{Introduction}
The channel coding theorem for a stationary and 
memoryless\footnote{
Throughout the paper, a stationary memoryless channel
without using entangled input states is 
simply referred to as a stationary memoryless channel; 
see Remark~\ref{remark:memoryless}.
}
(classical-)quantum channel has been established by combining 
the direct part shown by 
Holevo \cite{Holevo-QCTh} and 
Schumacher-Westmoreland 
\cite{Schumacher-Westmoreland} 
with the (weak) converse part which goes back to 
1970's works by 
Holevo\cite{Holevo-bounds,Holevo-bounds2}. 
This theorem is undoubtedly a landmark in the history 
of quantum information theory.  At the same time, however, we should 
not forget that stationary memoryless channels are 
not the only class of quantum channels.  It is indeed 
natural to think that many channels appearing in 
nature are neither stationary nor memoryless even 
in the approximate sense.  

In the classical information theory, a capacity formula 
for the most general setting 
was given by Verd\'{u} and Han 
\cite{Verdu-Han}, based on the so-called information-spectrum 
method \cite{Han_book}.  We show in this paper that a similar approach 
is applicable to yield some general formulas for the 
capacity of a classical-quantum channel (or in other words the 
classical-capacity of a quantum channel) and related 
notions.  

Let us take a brief look at the general feature of the 
information-spectrum method in the classical information theory.  
One of the main subjects of the information theory is 
to characterize asymptotic optimalities of various types of 
coding problems by entropy-like information quantities. 
In the information-spectrum method, a coding problem is 
treated in the most general setting, without assuming 
any structural assumptions such as the stationary memoryless 
property, and the asymptotic optimality is characterized 
by a limiting expression on information spectra (i.e., asymptotic 
behaviors of logarithmic likelihoods).  Since the asymptotic 
optimization of coding is essentially solved in 
this characterization, rewriting the 
information-spectrum quantity to an entropy-like quantity for 
a specific situation is mostly a direct consequence of 
a limiting theorem in the probability theory such as 
the law of large numbers, the Shannon-McMillan-Breiman theorem, 
ergodic theorems, large deviation theorems, etc. 
Such a framework brings not only generality 
but also transparency of mathematical arguments. 
Indeed, we are often led to simplification of a proof 
of an existing coding theorem by investigating it 
from the information-spectrum viewpoint. 

Turning to the quantum information theory, 
in spite of the recent remarkable 
progress of the field we often see that 
mathematical arguments to prove theorems 
are neither so 
transparent nor unified as in the classical theory. 
For instance, the original proof of the direct part of 
quantum channel coding theorem 
\cite{Holevo-QCTh, Schumacher-Westmoreland} 
is rather complicated so that it is not easy to 
grasp the essence of the argument(; see 
\cite{Winter} for a different proof).  
Extending the information-spectrum method to 
the quantum case is an attractive subject which 
brings a hope that proofs will be simplified 
and, more importantly, that 
both the optimality of coding systems and the limiting law
governing quantum stochastic situations 
will be provided with transparent and comprehensive understanding. 

In this paper, we pursue this subject for the quantum 
channel coding problem, whereby the quantum analogue of 
Verud\'{u}-Han's general formula is obtained.  In addition, 
the formula is applied to 
the stationary memoryless case to yield a new 
proof of the quantum channel coding theorem.  
It should be noted here that, in both of derivation 
of the general formula and application to the stationary 
memoryless case to get a nonasymptotic expression, 
there arise several mathematical difficulties to which the 
corresponding classical arguments are not immediately applicable. 
The difficulties in deriving the general formula 
are overcome by using the quantum Neyman-Pearson lemma 
\cite{Ho72,Hel, Nag-Hay:test}
and a novel operator inequality 
(Lemma~\ref{lemma:ineq_for_direct}), while those 
in rewriting the formula to the known form in 
the stationary memoryless case are coped with 
by invoking the asymptotic theory 
of hypothesis testing for two quantum states 
\cite{Hiai-Petz, Oga-Nag:test, Nag-Hay:test} 
(; see the references of \cite{Nag-Hay:test} for 
related results) as a kind of substitute of 
the weak law of large numbers. 
In particular, the inequality of Lemma~\ref{lemma:ineq_for_direct} 
is expected to play a key role in analyzing a measurement of 
the square root type in general; 
actually it drastically 
simplifies the original proof of 
\cite{Holevo-QCTh, Schumacher-Westmoreland} 
as mentioned in Remark~\ref{remark:simplify_HSW}. 

Historically, the present work is preceded by 
Ogawa's proof \cite{OD} of the direct part of 
the quantum channel coding theorem, with 
an improved and simplified version being 
found in \cite{ON1}, which 
was actually the first remarkable result of 
the information-spectrum approach to  
the quantum channel coding problem and elucidated 
the close relation between 
the channel coding and the hypothesis testing
in the quantum information theory; see 
Remark~\ref{remark:OD1} and Remark~\ref{remark:OD2}. 
In the present paper, we clarify this relation from a more general
viewpoint and make further developments
to establish the information-spectrum method in the quantum
channel coding theory.  
These attempts lead us to better understanding of 
the reason why the quantum relative entropy plays important roles
in both of these problems.

We should emphasize, however, that 
the present paper is not 
the final goal for the information-spectrum study of quantum channel capacity. 
Even though a general capacity formula has been given 
in terms of the quantum information spectrum, the way to apply it 
to the stationary memoryless case shown in this paper 
is not so straightforward as the classical counterpart.
Indeed, if our concern is restricted to proving the coding theorem for 
stationary memoryless channels, 
the information spectrum appears to be a kind of roundabout at present; 
see Remarks~\ref{remark:ch_from_hypo}, 
\ref{remark:ch_from_OH1} and \ref{remark:simplify_HSW}. 
In order to achieve the same level of simplicity and transparency as 
the classical information-spectrum method and to 
fulfill further the above-mentioned hope for the 
quantum information-spectrum method, 
we will need to have more theoretical tools 
to analyze the quantum information spectrum. 

The paper is organized as follows. In section~\ref{s2} 
the notion of general classical-quantum channels is introduced 
and the coding problem for it is formulated.  Section~\ref{s3} 
is devoted to asserting the main theorem, which gives the 
general capacity formula and the characterization of 
strong converse property of a general channel, while 
the proof is given in section~\ref{s5} based on 
some lemmas prepared in section~\ref{s4}. 
Stationary memoryless channels are treated in 
section~\ref{s6} and section~\ref{s8}, the latter 
of which considers cost constraint on inputs, 
while section~\ref{s7} is devoted to revisiting 
the decoder introduced by Holevo-Schumacher-Westmoreland 
in view of comparison to our decoder used to prove the general 
formulas.  Section~\ref{s9} gives some concluding remarks. 

\section{Capacity of general 
classical-quantum channels}\label{s2}

A quantum communication channel is generally composed of 
the following constructs; 
(separable) Hilbert 
spaces $\cH_1$ and $\cH_2$ which respectively represent 
the quantum systems of the sender's and the receiver's sides, 
a trace preserving CP (completely positive) map $\Gamma$ from the 
trace-class operators on $\cH_1$ to those on $\cH_2$ 
which describes the change of sent states, and a map 
$V:\cX\rightarrow\cS(\cH_1)$ which represents the 
modulator to set the input state to $V_x$ according to the value of 
the control variable $x\in\cX$.  When our concern is 
restricted to sending classical messages via the channel, 
however, only the composite map $\Gamma\circ V:\cX\rightarrow 
\cS(\cH_2)$ is relevant, and hence in the sequel we call a map 
$\ch:\cX\ni x\mapsto\charg{x}\in\cS(\cH)$ a classical-quantum channel 
or simply a channel.  Here $\cX$ is an arbitrary (finite or infinite) set 
and $\cH$ is an arbitrary Hilbert space.  This definition corresponds to 
the classical one in which a channel is represented by 
a conditional probability $W:(x,y)\mapsto W(y\,|\,x)$ or equivalently 
by a map $W:x\mapsto W_x= W(\,\cdot\,|\,x)$.  

\begin{remark} 
\rm  In many papers treating the capacity of 
quantum memoryless channels 
(e.g., \cite{Holevo-QCTh, Schumacher-Westmoreland, 
Holevo-bounds, Winter, Oga-Nag:channel}), 
only the case when $\cX$ is a finite set is considered. 
Even though the restriction to the finite case may be 
sufficient to understand the 
essence of most (but not all) mathematical arguments 
for proving the capacity theorem, there is no reason 
to restrict ourselves to the finite case from the 
standpoint that the capacity is the maximum reliable 
transmission rate of all possible communication 
systems for a given quantum channel.  Indeed, 
a particularly important infinite case is when 
$\cX=\cS(\cH_1)$ and $\ch$ is a trace-preserving 
CP map. 
\end{remark}

\begin{remark} \rm  
The term ``classical-quantum channel" has been provided 
with several  
different meanings in the literature (cf.\ \cite{Hol:quant-ph9809023}).  
The present 
definition is similar to that of \cite{Hol:RMP77}, although 
some measure-theoretic assumptions were made there on 
both the set $\cX$ and the mapping $x\mapsto\charg{x}$ 
to consider a channel in a general and unified 
operator-algebraic setting. 
\end{remark}

\begin{remark}
\rm 
As was pointed out in \cite{Fuj-Nag}, the capacity 
problem for a channel $\ch:\cX\rightarrow\cS(\cH)$ 
relies only on its range $\{\charg{x}\,|\,x\in\cX\}$, 
and we can adopt the alternative definition in which 
an arbitrary subset of $\cS(\cH)$ is called a channel. 
In other words, we can assume, if we wish, 
with no loss of generality that every  
$\ch$ appearing in the sequel is the identity map on a subset 
$\cX\subset\cS(\cH)$. 
The reason for treating a map $\ch$ instead of 
its range is mainly that it enables us to introduce 
more readable and natural notations. 
\end{remark}

For an arbitrary channel $\ch:\cX\rightarrow\cS(\cH)$,
we call a triple $(N, \enc, \dec)$ a {\em code for $\ch$} 
when it consists of a natural number (size) $N$, a 
mapping (encoding) $\enc : \{1, \ldots, N\} \rightarrow  
\cX$ and a POVM (decoding) $\dec=\{\decarg{i}\}_{i=1}^N$ on 
$\cH$ such that $\sum_i \decarg{i} \leq I$, 
where $I- \sum_i \decarg{i}$ corresponds to the failure of decoding, 
and denote 
the totality of such codes by $\cC(\ch)$.
For a code $\code = (N, \enc, \dec) \in \cC(\ch)$, 
the code size and the average error probability are 
represented as 
\begin{gather}
 |\code | \defeq N , \quad \mbox{and} \\
\Pe [\code] \defeq 
\frac{1}{N}\sum_{i=1}^N (1- \Tr [\charg{\encarg{i}} \decarg{i}]).
\end{gather}

Now let us proceed to the asymptotic setting. 
Suppose that we are given a sequence $\cHvec = \{\cHn\}_{n=1}^\infty$ 
of Hilbert spaces and a sequence $\chvec = \{\chn\}_{n=1}^\infty$ 
of channels $\chn :\cXn\rightarrow\cS(\cHn)$.  An important example 
is the stationary memoryless case 
when $\cHvec$ and $\chvec$ are 
defined from a Hilbert space $\cH$ and a channel $\ch :
\cX\rightarrow\cS(\cH)$ as 
$\cHn=\cHtensor$,  $\cXn=\cXtn$ and 
$\chnarg{\xn}=
\charg{x_1}\otimes\cdots\otimes\charg{x_n}$ 
for  $\xn=(x_1, \ldots , x_n)$, 
which will be treated in 
sections \ref{sec:memoryless} and \ref{sec:cost}. 
Except for those sections, however, we do not make any 
assumptions on the mutual 
relations among $\{\cHn\}$, $\{\cXn\}$ and $\{\chn\}$ for 
different $n$'s.  Such an extremely general setting is one of 
the main features of the information spectrum approach. 
The {\em capacity} of $\chvec$ is then defined as
\begin{align}
C(\chvec) \defeq 
\sup\,\{R\,|\, & \exists \codevec=\{\coden\}\in \cCvec (\chvec), 
\nonumber
\\
& \liminf_{n\rightarrow\infty}\frac{1}{n} \log |\coden | 
\geq R\;\;\mbox{and}\;\; 
\lim_{n\rightarrow\infty} \Pe[\coden] =0\,\}, 
\Label{def_C_general}
\end{align}
where $\cCvec (\chvec)$ denotes the totality 
of sequences of codes $\codevec =\{\coden\}_{n=1}^\infty$ such that 
$\coden\in \cC (\chn)$ for all $n$. 
We also introduce a `dual' of the capacity
\begin{align}
C^\dagger (\chvec) \defeq 
\inf\,\{R\,|\, & \forall \codevec=\{\coden\}\in \cCvec (\chvec), 
\nonumber
\\
&\liminf_{n\rightarrow\infty}\frac{1}{n} \log |\coden | 
\geq R\;\;\mbox{implies}\;\; 
\lim_{n\rightarrow\infty} \Pe[\coden] =1\,\} .
\Label{def_C+_general}
\end{align}
Note that $C(\chvec)\leq C^\dagger (\chvec) $ always 
holds. 
Following the terminology of classical information theory, 
we say that 
the {\em strong converse} holds for $\chvec$ when 
$C(\chvec) = C^\dagger (\chvec)$.

\section{Main results}\label{s3}

In this section we give general formulas for $C(\chvec)$ 
and $C^\dagger (\chvec)$ which are regarded as the quantum extensions 
of those for classical channel coding obtained by Verd\'{u} and 
Han \cite{Verdu-Han}.  The classical formula was given in terms 
of some information-spectrum-theoretic quantities, and we first 
need to introduce 
quantum analogues of these concepts along the 
line developed in \cite{Nag-Hay:test}. 

For a  self-adjoint trace-class operator $A$ 
with the spectral 
decomposition 
$A = \sum_{i} \lambda_i E_i$, where 
$\{\lambda_i\}$ are the eigenvalues and 
$\{E_i\}$ are the orthogonal 
projections onto the corresponding 
eigenspaces, we define 
\begin{align}
\spec{A\geq 0} \defeq \sum_{i:\lambda_i\geq 0} E_i 
\quad \mbox{and}\quad
\spec{A > 0} \defeq \sum_{i:\lambda_i > 0} E_i . 
\Label{12-13-1}
\end{align}
These are the orthogonal projections onto 
the direct sum of eigenspaces corresponding to 
nonnegative and positive eigenvalues, respectively. 
The projections $\spec{A\leq 0}$ and $\spec{ A < 0}$ 
are defined similarly. 

For any set $\cX$, let $\cP (\cX)$ be 
the totality of probability distributions on $\cX$ with 
finite supports.  That is, an element $P$ of $\cP (\cX)$ 
is a function $\cX\rightarrow[0,1]$ such that  
its support ${\rm supp} (P) \defeq \{x\,|\,P(x) >0\}$ 
is a finite set and that 
\[ \sum_{x\in\cX} P(x) = \sum_{x\in{\rm supp} (P)} P(x) 
=1 .\]
Let the totality of sequences $\Pvec = \{\Pn\}_{n=1}^\infty$ of 
$\Pn\in \cP (\cXn)$ be denoted by $\cPvec (\cXvec)$, and 
the totality of $\sigmavec = \{\sigman\}_{n=1}^\infty$ of 
$\sigman\in\cS (\cH)$ by $\cSvec (\cHvec)$. 
Given $\Pvec\in\cPvec(\cXvec)$ and $\sigmavec
\in\cSvec(\cHvec)$, let 
\begin{align*}
\supJ (\Pvec, \sigmavec, \chvec)  
&\defeq 
\inf\,\left\{a \left|\, 
\lim_{n\rightarrow\infty} \sum_{\xn\in\cXn} 
\Pn(\xn) \Tr \left[ \chnarg{\xn} \spec{\chnarg{\xn} - e^{na}\sigman >0}
\right] = 0\right.\right\}, 
\\
\infJ (\Pvec, \sigmavec, \chvec) 
&\defeq 
\sup\, \left\{a\, \left|\, 
\lim_{n\rightarrow\infty} \sum_{\xn\in\cXn} 
\Pn(\xn) \Tr \left[\chnarg{\xn} \spec{\chnarg{\xn} - e^{na}\sigman >0}
\right] = 1\right.\right\}, 
\end{align*}
and 
\begin{align*}
\supI (\Pvec, \chvec) 
&\defeq 
\supJ (\Pvec, \chvecarg{\Pvec}, \chvec), \\
\infI (\Pvec, \chvec) 
&\defeq 
\infJ (\Pvec, \chvecarg{\Pvec}, \chvec), 
\end{align*}
where $\chvecarg{\Pvec}$ denotes the sequence 
$\{\chnarg{\Pn} \}_{n=1}^\infty$ of 
\begin{align}
\Label{def:W_Pn}
 \chnarg{\Pn} \defeq 
\sum_{\xn\in\cXn} \Pn (\xn) \, \chnarg{\xn} 
\; \in \cS (\cHn) .
\end{align}
Note that $\supI (\Pvec, \chvec) $ and 
$\infI (\Pvec, \chvec) $ are quantum analogues 
of the spectral sup- and inf-information rates (\cite{Verdu-Han}):
\begin{align*}
\supI (\bX\,;\, \bY) &\defeq \mbox{p-}\limsup_{n\rightarrow\infty} 
\frac{1}{n} \log \frac{\Wn (\Yn\,|\,\Xn)}{P_{\Yn}(\Yn)}, \\
\infI (\bX\,;\, \bY) &\defeq \mbox{p-}\liminf_{n\rightarrow\infty} 
\frac{1}{n} \log \frac{\Wn (\Yn\,|\,\Xn)}{P_{\Yn}(\Yn)},
\end{align*}
where $\bY=\{\Yn\}$ is supposed to be the sequence of 
random variables obtained as the outputs of channels 
$\bW=\{\Wn\}$ for a sequence of input random variables $\bX=\{\Xn\}$.

\begin{remark}
\rm 
The projection $\spec{\chnarg{\xn} - e^{na}\sigman >0}$ in 
the definitions above can be 
replaced with $\spec{\chnarg{\xn} - e^{na}\sigman \geq 0}$ or, 
more generally, 
with an arbitrary self-adjoint operator $S$ satisfying 
\[
\spec{\chnarg{\xn} - e^{na}\sigman >0} \leq 
S \leq \spec{\chnarg{\xn} - e^{na}\sigman \geq 0}. 
\]
 This ambiguity 
does not influence the definitions of the above quantities; 
see \cite{Nag-Hay:test}. 
\end{remark}

Now we have the following theorem.

\begin{theorem}
\Label{thm:main}
\begin{align}
C(\chvec) &= 
\max_{\Pvec\in\cPvec(\cXvec)} \; \infI (\Pvec, \chvec) 
\Label{generalC.1}
\\ &=
\max_{\Pvec\in\cPvec(\cXvec)} \; \min_{\sigmavec\in\cSvec(\cHvec)} \; 
\infJ (\Pvec, \sigmavec, \chvec) ,
\Label{generalC.2}
\end{align}
and
\begin{align}
C^\dagger (\chvec) &= 
\max_{\Pvec\in\cPvec(\cXvec)} \; \supI (\Pvec, \chvec) 
\Label{generalC+.1} 
\\ &=
\max_{\Pvec\in\cPvec(\cXvec)} \; \min_{\sigmavec\in\cSvec(\cHvec)} \; 
\supJ (\Pvec, \sigmavec, \chvec) .
\Label{generalC+.2}
\end{align}
\end{theorem}

\begin{remark}
\rm The formula obtained by Verd\'{u} and Han 
\cite{Verdu-Han} for a sequence of classical channels $\bW = 
\{\Wn\}_{n=1}^\infty$ is 
\begin{align}
\label{classical_C}
C(\bW) = \sup_{\bX}\infI (\bX\,;\, \bY), 
\end{align}
where the supremum  is taken over all possible 
input sequences $\bX =\{\Xn\}$, and $\bY =\{\Yn\}$ 
denotes the output sequences corresponding to 
$\bX$. 
In addition, they showed that the strong converse 
holds for $\bW$ if and only if 
$\sup_{\bX}\infI (\bX\,;\, \bY) = \sup_{\bX}\supI (\bX\,;\, \bY)$.  
In the process of proving this, they have essentially shown that 
\begin{align}
\label{classical_C+}
C^\dagger (\bW) = \sup_{\bX}\supI (\bX\,;\, \bY) ,
\end{align}
even though $C^\dagger (\bW) $ does not explicitly appear in 
that paper. Note that the supremums in these expressions 
can be replaced with maximums 
(see Remark \ref{remark:sup=max_inf=min} below), 
and our expressions (\ref{generalC.1}) and (\ref{generalC+.1}) are the 
quantum extensions of (\ref{classical_C}) and (\ref{classical_C+}). 
\end{remark}

\begin{remark}
\label{remark:J>=I}
\rm
In the classical case, let 
\begin{align*}
\supJ (\bX, \bf{\hat{Y}}, \bW) &\defeq \mbox{p-}\limsup_{n\rightarrow\infty} 
\frac{1}{n} \log \frac{\Wn (\Yn\,|\,\Xn)}{P_{\hat{Y}^{(n)}}(\Yn)}, \\
\infJ (\bX, \bf{\hat{Y}}, \bW) &\defeq \mbox{p-}\liminf_{n\rightarrow\infty} 
\frac{1}{n} \log \frac{\Wn (\Yn\,|\,\Xn)}{P_{\hat{Y}^{(n)}}(\Yn)}, 
\end{align*}
where $\hat{Y}^{(n)}$ is an arbitrary random variable with a 
probability distribution $P_{\hat{Y}^{(n)}}$ taking values in a common  
set with $\Yn$.  Then we have
\begin{align*}
\supJ (\bX, \hat{\bf Y}, \bW) &\geq \mbox{p-}\limsup_{n\rightarrow\infty} 
\frac{1}{n} \log \frac{\Wn (\Yn\,|\,\Xn)}{P_{\Yn}(\Yn)}
+
\mbox{p-}\liminf_{n\rightarrow\infty} 
\frac{1}{n} \log \frac{P_{\Yn}(\Yn)}{P_{\hat{Y}^{(n)}}(\Yn)}\\
&= 
\supI (\bX\,;\,\bY) + \infD (\bY\,\|\,\bf{\hat{Y}}), \\
\infJ (\bX, \hat{\bf Y}, \bW) &\geq \mbox{p-}\liminf_{n\rightarrow\infty} 
\frac{1}{n} \log \frac{\Wn (\Yn\,|\,\Xn)}{P_{\Yn}(\Yn)}
+
\mbox{p-}\liminf_{n\rightarrow\infty} 
\frac{1}{n} \log \frac{P_{\Yn}(\Yn)}{P_{\hat{Y}^{(n)}}(\Yn)}\\
&= 
\infI (\bX\,;\,\bY) + \infD (\bY\,\|\,\bf{\hat{Y}}),
\end{align*}
where 
$\infD (\bY\,\|\,\bf{\hat{Y}})$ is the spectral 
inf-divergence rate \cite{Han_book} between $\bY$ and $\bf{\hat{Y}}$. 
Since $\infD (\bY\,\|\,\bf{\hat{Y}}) \geq 0$ always holds, 
we have 
\begin{align}
\supI (\bX\,;\,\bY)
&= 
\min_{\hat{\bf Y}} \supJ (\bX, \hat{\bf Y}, \bW), 
\quad\mbox{and}
\\
\infI (\bX\,;\,\bY)
&= 
\min_{\hat{\bf Y}} \infJ (\bX, \hat{\bf Y}, \bW),
\end{align}
which yield similar expressions to (\ref{generalC.2}) and (\ref{generalC+.2}) 
from (\ref{classical_C}) and (\ref{classical_C+}). 
In the quantum case, on the other hand, it is not clear 
whether the corresponding equations 
$\supI (\Pvec, \chvec) = \min_{\sigmavec}\supJ (\Pvec, \sigmavec, \chvec)$ 
and 
$\infI (\Pvec, \chvec) =\min_{\sigmavec}\infJ (\Pvec, \sigmavec, \chvec)$ 
generally hold.  
Nevertheless the expressions for $C(\chvec)$ and $C^\dagger(\chvec)$ 
in 
Theorem~\ref{thm:main} always hold. 
\end{remark}

\begin{remark} 
\label{remark:sup=max_inf=min}
\rm 
If a classical or quantum information-spectrum quantity 
includes a sequence of variables, the supremum (infimum, resp.)
 (e.g.\ (\ref{classical_C}), (\ref{classical_C+}) ) with respect to 
the variables can always be replaced with the maximum (miminum) 
due to the following lemma.  Thus we do not need to care about the attainability 
of such a supremum (infimum).  
\end{remark}

\begin{lemma}
\label{lemma:sup_inf}
Suppose that we are given a sequence 
$\{\cFn\}_{n=1}^\infty$, where each $\cFn$ 
is a nonempty set consisting 
of monotonically nondecreasing functions defined on  
$\bR$, and let $\cFvec$ denote the totality of 
sequences $\fvec =\{\fn\}_{n=1}^\infty$ 
of functions $\fn\in\cFn$; in other words, $\cFvec$ is the 
direct product $\prod_{n=1}^\infty\cFn$ of $\{\cFn\}$.  
For each $\fvec\in\cFvec$ and $x\in\bR$, let 
\begin{align*}
[\fvec]^-_x &\defeq 
\sup\,\{a\,|\,\limsup_{n\rightarrow\infty} \fn (a) \leq x\} \;\in \bR\cup\{\infty, -\infty\}, \\
[\fvec]_x^+ &\defeq 
\inf\,\{a\,|\,\liminf_{n\rightarrow\infty} \fn (a) \geq x\}  \;\in \bR\cup\{\infty, -\infty\}.
\end{align*}
Then the supremums and infimums of 
\[ \sup_{\fvec}\, [\fvec]^-_x, \;\; 
\sup_{\fvec}\, [\fvec]^+_x, \;\; 
\inf_{\fvec}\, [\fvec]^-_x \quad\mbox{and}\quad
\inf_{\fvec}\, [\fvec]^+_x
\]
are always attainable in $\cFvec$. 
\end{lemma}

\begin{proof} See Appendix I. 
\end{proof}

In the situation of Thorem~\ref{thm:main}, for instance, the lemma is 
applied to sequences of functions $\fvec =\{\fn\}_{n=1}^\infty$ of the form
\[ f_n (a) = \sum_{\xn\in\cXn} 
\Pn(\xn) \Tr \left[\chnarg{\xn} \spec{\chnarg{\xn} - e^{na}\sigman \leq 0}
\right], 
\]
for which we have
$[\fvec]^-_0 = \infJ (\Pvec, \sigmavec, \chvec)$ and 
$[\fvec]^+_1 = \supJ (\Pvec, \sigmavec, \chvec)$.  
Note that the monotonicity of these functions follows
from an argument in section 3 of \cite{Nag-Hay:test}. 

\section{
Lemmas for proving 
Theorem \ref{thm:main}}\label{s4}

We need three lemmas.  The first one 
is the key operator inequality to prove the second, 
while the 
second and third lemmas are directly used to prove 
the theorem.  Throughout this paper the generalized inverse of 
a nonnegative operator $A$ is simply denoted by $A^{-1}$; 
i.e., $A^{-1}$ is the nonnegative operator such that 
$A A^{-1}=A^{-1}A= P_A= P_{A^{-1}}$ where $P_A$ and $P_{A^{-1}}$ denote 
the orthogonal projections onto the ranges of $A$ and $A^{-1}$. 

\begin{lemma}
\Label{lemma:ineq_for_direct}
For any  positive number $c$ and any 
operators 
$0 \le S \le I$ and $T\ge 0$, we have
\begin{align}
I- \sqrt{S+T}^{-1} S \sqrt{S+T}^{-1}
\leq 
(1+c)\, (I -S) + (2+c+c^{-1})\;T. \Label{h1}
\end{align}
\end{lemma}

\begin{proof}
Let $P$ be the orthogonal projection onto the range of $S+T$. 
Then $P$ commutes both $S$ and $T$, and hence  it is enough to prove 
\begin{align*}
&P \left[I- \sqrt{S+T}^{-1} S \sqrt{S+T}^{-1}\right] P
\leq 
P \left[(1+c)\, (I -S) + (2+c+c^{-1})\,T\right] P, \quad\mbox{and} \\
&P^\bot\left[I- \sqrt{S+T}^{-1} S \sqrt{S+T}^{-1}\right] P^\bot
 \leq 
P^\bot \left[ (1+c)\,(I -S) + (2+c+c^{-1})\,T\right] P^\bot  ,
\end{align*}
where $P^\bot=I-P$. 
Since $P^\bot S=P^\bot T=P^\bot \sqrt{S+T}^{-1}=0$,
the second inequality is trivial. 
Thus, we have only to show the first one or, equivalently, 
to show (\ref{h1}) 
in the case when the range of $S+T$ is ${\cal H}$. 
Substituting $A=\sqrt{T}$ and $B=\sqrt{T}( \sqrt{S+T}^{-1} - I)$ 
into the general operator inequality 
$A^*B + B^* A \le c^{-1}\, A^* A + c\, B^* B$, which follows from 
$(A-c B)^* (A-cB)\geq 0$, we have
\begin{align}
& T ( \sqrt{S+T}^{-1} - I)
+ ( \sqrt{S+T}^{-1} - I)T \nonumber \\
\le & \; 
c^{-1}\,T +  c\,( \sqrt{S+T}^{-1} - I)T ( \sqrt{S+T}^{-1} - I). \Label{h2}
\end{align}
In addition, 
since the function $f(x) = \sqrt{x}$ is an operator 
monotone function and $0\leq S\leq I$, 
we have 
\begin{align}
\sqrt{S+T}  \ge \sqrt{S} \ge S
. \Label{h3}
\end{align}
Now the desired inequality is proved as follows:
\begin{align*}
&I- \sqrt{S+T}^{-1} S \sqrt{S+T}^{-1}  
= \sqrt{S+T}^{-1} T \sqrt{S+T}^{-1}\\
=&  T + T ( \sqrt{S+T}^{-1} - I)
+ ( \sqrt{S+T}^{-1} - I)T 
+ ( \sqrt{S+T}^{-1} - I)T ( \sqrt{S+T}^{-1} - I)\\
\le& 
(1+c^{-1})\, T+ (1+c)\, ( \sqrt{S+T}^{-1} - I)T ( \sqrt{S+T}^{-1} - I)\\
\le&
(1+c^{-1})\, T+  (1+c)\, ( \sqrt{S+T}^{-1} - I)(S+T) ( \sqrt{S+T}^{-1} - I)\\
= &
 (1+c^{-1})\,T+  (1+c)\, ( I + S +T - 2 \sqrt{S+T}) \\
\le&   (1+c^{-1})\,T+ (1+c)\, ( I + S +T - 2 S) \\
=& (1+c)\, ( I - S) + (2+c+c^{-1})\,T ,
\end{align*}
where the first inequality follows from (\ref{h2})
and the third inequality follows from (\ref{h3}).
\end{proof}

\begin{lemma}
\Label{lemma:direct}
For any $n\in\bN$, $a\in\bR$, $N\in\bN$,  
$\Pn\in\cP (\cXn)$ and $c>0$, 
there exists a code $\coden\in\cC (\chn)$ such that 
$|\coden | =N$ and 
\begin{align}
\nonumber
\Pe [\coden] \leq 
&\;
(1+c) \sum_{\xn\in\cXn} \Pn(\xn) 
\Tr \left[ \chnarg{\xn} \spec{\chnarg{\xn} - e^{na}\chnarg{\Pn} \leq 0} \right] 
\\
&+ (2+c+c^{-1}) e^{-n a} N ,
\label{ineq:direct}
\end{align}
where $\chnarg{\Pn}$ is defined by (\ref{def:W_Pn}).
\end{lemma}
\begin{proof}
We prove the lemma by a random coding method.
Given $n$, $a$, $N$, $\Pn$ and an encoder $\encn :\{1, \ldots, N\}\rightarrow\cXn$, 
define the decoding POVM $\decn=\{\decnarg{i}\}_{i=1}^N$ by
\begin{align}
\label{def_Xni}
\decnarg{i} &\defeq 
\left(\sum_{j=1}^N \pi_j\right)^{-\frac{1}{2}}
\pi_i
\left(\sum_{j=1} ^N
\pi_j\right)^{-\frac{1}{2}} ,
\end{align}
where 
\begin{equation}
\label{def_pii}
\pi_i \defeq\spec{\chnarg{\encnarg{i}} - e^{na} \chnarg{\Pn} \,> 0}.
\end{equation}
Denoting the average error probability 
$\Pe[\coden]$ of the code $\coden =(N, \encn, \decn)$ by
$\Pe[\encn]$, we have
\begin{align}
\Pe[\encn]
&=
\frac{1}{N}\sum_{i=1}^N
\Tr \left [ \chnarg{\encnarg{i}} 
\left( I - \decnarg{i} \right) \right]
\nonumber \\
& \le
\frac{1}{N}\sum_{i=1}^N
\Tr \left [\chnarg{\encnarg{i}} 
\left( (1+c)\,(I - \pi_i)
+ (2+c+c^{-1})\,\sum_{j \neq i}
\pi_j
\right)
\right ],
\Label{h4}
\end{align}
which follows from Lemma \ref{lemma:ineq_for_direct}. 
Now suppose that an encoder $\encn$ is randomly generated according to the 
 probability distribution $\Pnrc(\encn) = 
\Pn(\encnarg{1})\cdots \Pn(\encnarg{N})$.  
The expectation of $\Pe[\encn]$ under
$\Pnrc$ is then bounded from above as 
\begin{align}
\nonumber 
\Enrc \Pe[\encn] 
\le &
\Enrc
\frac{1+c}{N}\sum_{i=1}^N
\Tr \left[\chnarg{\encnarg{i}} 
\spec{\chnarg{\encnarg{i}} - e^{na} \chnarg{\Pn} \le 0}\right] \\
\nonumber 
&+ 
\Enrc 
\frac{2+c+c^{-1}}{N}\sum_{i=1}^N \sum_{j \neq i}
\Tr \left[\chnarg{\encnarg{i}} \spec{\chnarg{\encnarg{j}} - e^{na} \chnarg{\Pn} \,> 0}
\right] \\
\nonumber 
= &
(1+c) \sum_{\xn}\Pn(\xn) 
\Tr\left[ \chnarg{\xn}\spec{\chnarg{\xn} - e^{na} \chnarg{\Pn} \le 0}\right] \\
\Label{h5}
& +
(2+c+c^{-1}) N
\sum_{\xn} \Pn(\xn)  
\Tr \left[\chnarg{\Pn}\spec{\chnarg{\xn}-  e^{na} \chnarg{\Pn}\,> 0} \right] .
\end{align}
Substituting $A=\chnarg{\xn}-  e^{na} \chnarg{\Pn}$ into 
$\Tr[A\spec{A>0}]\geq 0$, 
the second term of (\ref{h5}) is further evaluated by
\begin{align*}
&
\sum_{\xn} \Pn(\xn)  
\Tr \left[\chnarg{\Pn}\spec{\chnarg{\xn}-  e^{na} \chnarg{\Pn}\,> 0} \right] \\
\le &
 e^{-na}
\sum_{\xn} \Pn(\xn)  
\Tr \left[ \chnarg{\xn} 
\spec{\chnarg{\xn}-  e^{na} \chnarg{\Pn}\,> 0} \right] \\
\le &
e^{-na} . 
\end{align*}
Thus the existence of $\encn$ for which the code 
$\coden = (N, \encn, \decn)$ satisfies (\ref{ineq:direct}) has been 
proved. 
\end{proof}

\begin{remark}
\label{remark:ineq_direct_classical}
\rm
In deriving the direct part of the general capacity formula 
for classical channels, Verd\'{u} and Han \cite{Verdu-Han} 
invoked the so-called Feinstein's lemma (Theorem~1 in \cite{Verdu-Han}; 
see the next remark) 
which ensures the existence of a code satisfying
\begin{equation}
\Pe [\coden] \leq 
\;
{\rm Prob}\left\{
\frac{1}{n} \log 
\frac{\Wn (\Yn\,|\,\Xn)}{P_{\Yn}(\Yn)}
\leq a
\right\}
+ e^{-n a} N .
\label{ineq:direct_classical}
\end{equation}
Lemma~\ref{lemma:direct} above can be regarded as a quantum 
analogue of Feinstein's lemma, although the coefficients there 
are a bit larger.  
\end{remark}

\begin{remark}
\label{remark:OD1}
\rm
Historically, it seems that Shannon \cite{Shannon57}  was the first 
to explicitly formulate the inequality (\ref{ineq:direct_classical}).  
He used a random coding argument to prove that there exists a code whose 
{\em average} error probability satisfies (\ref{ineq:direct_classical}). 
On the other hand, Blackwell et al.\ \cite{BlaBreTho59} showed that 
the same inequality is also satisfiable for the {\em maximum} error probability.
They proved this by refining Feinstein's non-random packing argument, which 
is well known to have been used in 
the first rigorous proof of the coding theorem for 
discrete memoryless channels \cite{Feinstein54}. 
This course of things makes some people to call the 
theorem concerning (\ref{ineq:direct_classical}) 
``Feinstein's lemma", 
sometimes only for the maximum error probability and sometimes 
for both criteria (cf.\ \cite{Verdu-Han}).  
We note that the original proof of Feinstein does not 
yield the general capacity formula, and the 
refinement mede by Blackwell et al.\ is 
essential in this respect. 
Our Lemma~\ref{lemma:direct} corresponds to Shannon's one, 
while 
an attempt toward a quantum extension of the result of 
Blackwell et al.\ has been made  in \cite{OD, ON1}. 
The result obtained there is unfortunately not general enough to prove 
the direct part of the general formula (\ref{generalC.1}), but 
is of a particular interest itself; see Remark~\ref{remark:OD2} 
below. 
\end{remark}

\begin{remark}
\rm
Letting $A\defeq\sum_{\xn\in\cXn} \Pn(\xn) 
\Tr \left[ \chnarg{\xn} \spec{\chnarg{\xn} - e^{na}\chnarg{\Pn} \leq 0} \right]
$
and 
$B\defeq e^{-na} N$, 
the RHS of (\ref{ineq:direct}) is minimized at 
$c=\sqrt{\frac{B}{A+B}}$, which 
proves the existence of a code satisfying 
\[
\Pe [\coden] \leq A + 2B + 2\sqrt{B(A+B)} .
\]
\end{remark}

\begin{lemma}
\Label{lemma:converse}
For any $n\in\bN$ 
and any code $\coden\in\cC (\chn)$ with 
$|\coden | =N$, there exists a probability distribution $\Pn\in\cP (\cXn)$ 
such that for any $a\in\bR$ and $\sigman\in\cS (\cHn)$ 
\begin{align}
\label{ineq:converse}
\Pe [\coden] \geq \sum_{\xn\in\cXn} \Pn (\xn) 
\Tr \left[ \chnarg{\xn} \spec{\chnarg{\xn} - e^{na}\sigman \leq 0} \right]
- \frac{e^{na}}{N} .
\end{align}
\end{lemma}

\begin{proof}
Remember that for any operators $A\geq 0$ and 
$0\leq T\leq I$, 
\begin{equation}
\label{Neyman-Pearson}
\Tr\left[
A T
\right] \leq
\Tr\left[
A\spec{A>0}
\right],
\end{equation}
which is the essence of the quantum Neyman-Pearson lemma 
\cite{Ho72,Hel, Nag-Hay:test}.  Then we see that
for any code $\coden =(N,\encn, \decn)$, 
\begin{align*}
\Tr\left[
\left(\chnarg{\encnarg{i}} -e^{na} \sigman\right) 
\decnarg{i}\right]
&\leq 
\Tr\left[ 
\left(\chnarg{\encnarg{i}} -e^{na} \sigman\right) 
\spec{\chnarg{\encnarg{i}} -e^{na} \sigman\,>0}
\right] \\
& \leq 
\Tr\left[ 
\chnarg{\encnarg{i}} 
\spec{\chnarg{\encnarg{i}} -e^{na} \sigman\,>0}
\right] .
\end{align*}
This is rewritten as
\begin{align*}
\Tr\left[
\chnarg{\encnarg{i}} \left(I-\decnarg{i}\right) \right] 
\geq 
\Tr\left[ 
\chnarg{\encnarg{i}} 
\spec{\chnarg{\encnarg{i}} -e^{na} \sigman\leq 0}
\right]
- e^{na} \Tr\left[\sigman\decnarg{i}\right] ,
\end{align*} 
and hence we have
\begin{align*}
\Pe[\coden ] 
&\geq 
\frac{1}{N} \sum_{i=1}^N 
\Tr\left[ 
\chnarg{\encnarg{i}} 
\spec{\chnarg{\encnarg{i}} -e^{na} \sigman\leq 0}
\right] - 
\frac{e^{na}}{N}\sum_{i=1}^N \Tr\left[\sigman\decnarg{i}\right] \\
&\geq 
\frac{1}{N} \sum_{i=1}^N 
\Tr\left[ 
\chnarg{\encnarg{i}} 
\spec{\chnarg{\encnarg{i}} -e^{na} \sigman\leq 0}
\right] - 
\frac{e^{na}}{N}.
\end{align*}
We thus have  (\ref{ineq:converse}) by letting $\Pn$ 
be the empirical distribution for the $N$ points 
$(\encnarg{1}, \ldots, \encnarg{N})$. 
\end{proof}

\begin{remark}
\rm Lemma~\ref{lemma:converse} 
in the case of $\sigman=\chnarg{\Pn}$ 
is just the 
quantum analogue of Theorem~4 in 
\cite{Verdu-Han} which evaluates 
the error probability of a code as
\begin{equation}
\Pe [\coden] \geq 
\;
{\rm Prob}\left\{
\frac{1}{n} \log 
\frac{\Wn (\Yn\,|\,\Xn)}{P_{\Yn}(\Yn)}
\leq a
\right\}
- \frac{e^{n a}}{N} .
\label{ineq:converse_classical}
\end{equation}
Our results might seem to be still incomplete 
in comparison with 
the beautiful duality between 
(\ref{ineq:direct_classical}) and (\ref{ineq:converse_classical}).
\end{remark}

\section{Proof of Theorem \ref{thm:main}}
\label{s5}

Now Theorem \ref{thm:main} is proved as follows. 
We first show the inequality
\begin{equation}
C(\chvec) \geq 
\max_{\Pvec\in\cPvec(\cXvec)} \; \infI (\Pvec, \chvec). \Label{h9}
\end{equation}
Here 
we can assume that the RHS is strictly positive since otherwise the 
inequality is trivial.  Suppose that we are given 
a sequence $\Pvec =\{\Pn\}\in\cPvec(\cXvec)$ and a 
number $R$ such that $0<R <\infI (\Pvec,\chvec)$.  
Setting $N=\lceil e^{nR}\rceil$ in 
Lemma \ref{lemma:direct}, 
it follows that for each real number $a$ and $c>0$ 
there exists a sequence of codes $\codevec =\{\coden\}\in
\cCvec(\chvec)$ such that  
$|\coden | = \lceil e^{nR}\rceil$ and 
\begin{align}
\nonumber 
\Pe [\coden] \le
&
(1+c) \sum_{\xn\in\cXn} \Pn (\xn) 
\Tr 
\left[ \chnarg{\xn} 
\spec{\chnarg{\xn} - e^{na}\chnarg{\Pn} \leq 0} 
\right] \\
&
+(2+c+c^{-1}) \, e^{-na} \lceil e^{nR}\rceil 
\label{ineq:direct_var}
\end{align}
for every $n$. 
Recalling the definition of $\infI (\Pvec,\chvec)$, we see that 
the first term of the RHS goes to $0$ as $n\rightarrow\infty$ 
for any $a<\infI (\Pvec,\chvec)$, while the second term goes to $0$ 
for any $a>R$.  Hence, letting $a$ lie in $R <a<\infI (\Pvec,\chvec)$, 
the existence of a $\codevec$ satisfying 
$\liminf_{n\rightarrow\infty}\frac{1}{N}\log \coden | \geq R$ and 
$\lim_{n\rightarrow\infty}\Pe[\coden]=0$ is shown. 
This implies that $R\leq C(\chvec)$ for any $0<R< \infI (\Pvec,\chvec)$, 
and completes the proof of (\ref{h9}).  

Next we prove 
\begin{equation}
C^\dagger (\chvec) \geq 
\max_{\Pvec\in\cPvec(\cXvec)} \; \supI (\Pvec, \chvec). 
\label{h10}
\end{equation}
We can assume that $C^\dagger (\chvec) <\infty$ since 
otherwise the inequality is trivial.  Let $R$ be 
an arbitrary number greater than $C^\dagger (\chvec)$.  Then 
for each $a$ and $c>0$ there exists a sequence of codes 
$\codevec =\{\coden\}\in\cCvec(\chvec)$ such that  
$|\coden | = \lceil e^{nR}\rceil$ and  (\ref{ineq:direct_var}) 
holds for every $n$.  From 
$\lim_{n\rightarrow\infty}\frac{1}{n}\log |\coden | 
= R > C^\dagger (\chvec) $,  $\Pe[\coden]$ must go to
$1$ as $n\rightarrow\infty$, and therefore (\ref{ineq:direct_var}) 
yields that for any $a>R$ 
\[
1\leq (1+c) \liminf_{n\rightarrow\infty} 
\sum_{\xn\in\cXn} \Pn (\xn) 
\Tr 
\left[ \chnarg{\xn} 
\spec{\chnarg{\xn} - e^{na}\chnarg{\Pn} \leq 0} 
\right] .
\]
Since $c>0$ is arbitrary, 
$
\sum_{\xn\in\cXn} \Pn (\xn) 
\Tr 
\left[ \chnarg{\xn} 
\spec{\chnarg{\xn} - e^{na}\chnarg{\Pn} \leq 0} \right]
$ converges to $1$ 
and hence 
$a \geq \supI (\Pvec,\chvec)$. 
We thus have $a \geq \supI (\Pvec,\chvec)$ 
for 
$\forall a > \forall R > C^\dagger (\chvec) $, 
and (\ref{h10}) has been proved. 

Let us proceed to prove the converse inequality
\begin{equation}
C(\chvec) \leq 
\max_{\Pvec\in\cPvec(\cXvec)} \; 
\min_{\sigmavec\in\cSvec(\cHvec)} \; 
\infJ (\Pvec, \sigmavec, \chvec).  \label{h11} 
\end{equation}
Let $R < C(\chvec)$. Then there exists 
a sequence of codes $\codevec=\{\coden\}\in\cCvec
(\chvec)$  satisfying
\[  \liminf_{n \to \infty}\frac{1}{n} \log |\coden | > R
\quad\mbox{and}\quad
\lim_{n \to \infty} \Pe [\coden]= 0.
\] 
\par\noindent From Lemma \ref{lemma:converse}, 
there exists a 
$\Pvec= \{ \Pn \}$ such that 
for any $n \in \bN$ and 
$\sigmavec =\{  \sigman \}$, 
\begin{align*}
\sum_{\xn\in\cXn} \Pn (\xn) 
\Tr \left[ \chnarg{\xn} 
\spec{\chnarg{\xn} - e^{nR}\sigman \leq 0} \right]
&\le \Pe [\coden] + \frac{e^{nR}}{|\coden |} \\
&\to 0
\;\;\mbox{as}\;\; n\to\infty.
\end{align*}
This implies that $R \le \min_{\sigmavec}\; \infJ (\Pvec, \sigmavec, \chvec) $ 
for some $\Pvec$. 
Therefore, we have 
\[ R \leq 
\max_{\Pvec\in\cPvec(\chvec)} \; 
\min_{\sigmavec} \; 
\infJ (\Pvec, \sigmavec, \chvec)
\]
for any $R<C(\chvec)$, 
and 
 (\ref{h11}) has been proved. 
Similarly, we can prove
\begin{equation}
C^\dagger (\chvec) \leq 
\max_{\Pvec\in\cPvec(\cXvec)} \; 
\min_{\sigmavec\in\cSvec(\cHvec)} \; 
\supJ (\Pvec, \sigmavec, \chvec).  \label{h12}
\end{equation}
The remaining parts 
\[
\max_{\Pvec\in\cPvec(\cXvec)} \; \infI (\Pvec, \chvec) 
\geq 
\max_{\Pvec\in\cPvec(\cXvec)} \; 
\min_{\sigmavec\in\cSvec(\cHvec)} \; 
\infJ (\Pvec, \sigmavec, \chvec) 
\]
and 
\[
\max_{\Pvec\in\cPvec(\cXvec)} \; \supI (\Pvec, \chvec) 
\geq 
\max_{\Pvec\in\cPvec(\cXvec)} \; 
\min_{\sigmavec\in\cSvec(\cHvec)} \; 
\supJ (\Pvec, \sigmavec, \chvec) 
\]
are obvious from the definitions. 
	
\section{Stationary memoryless case}\label{s6}
\Label{sec:memoryless} 

In this section we demonstrate how the general 
formulas given in Theorem \ref{thm:main} leads to 
the following coding theorem for 
stationary memoryless channels.  

\begin{theorem}
Let $\ch:\cX\rightarrow\cS(\cH)$ be an arbitrary 
channel and consider its stationary memoryless 
extension:
\begin{align}
\nonumber 
&\cHn=\cHtensor, \quad \cXn=\cXtn, \quad
\mbox{and}\\
\label{cond:memoryless}
&\chnarg{\xn}=
\charg{x_1}\otimes\cdots\otimes\charg{x_n}\quad
\mbox{for}\quad \xn=(x_1, \ldots , x_n). 
\end{align}
Then the capacity of $\chvec=\{\chn\}$ is given by
\begin{align}
\Label{capacity_memoryless.1} 
C(\chvec) = 
\sup_{P\in\cP (\cX)} I(P,\ch),
\end{align}
where
\[
I(P,\ch)\defeq \sum_{x\in\cX} P(x) D(\charg{x}\,\|\,\charg{P}) 
\]
with $D(\rho\,\|\,\sigma)\defeq 
\Tr[\rho (\log\rho - \log\sigma)]$ 
being the quantum relative entropy. 
Furthermore, if $\dim\cH <\infty$,  
then the strong converse holds: $C^\dagger (\chvec) = C(\chvec)$. 
\end{theorem}

\begin{remark}
\label{remark:dimH}
\rm 
The proof of the strong converse given below relies 
essentially on the compactness of 
the closure of the range $\Delta=\{\charg{x}\,|\,x\in\cX\}$, which 
follows from the finiteness of $\dim\cH$. 
The argument is immediately extended to a certain class of 
channels with $\dim\cH=\infty$ including 
the case when $\cX$ is a finite set, whereas the general condition 
for the strong converse in the infinite-dimensional case is 
yet to be studied. 
\end{remark}

\begin{remark}
\label{remark:memoryless}
\rm
Let $\Gamma$ be a trace-preserving CP map from 
the trace-class operators on ${\cH}_1$ to those on 
${\cH}_2$.  When considering $\Gamma$ as a classical-quantum
channel 
$\ch:\cX\rightarrow\cS({\cH}_2)$ with $\cX=\cS ({\cH}_1)$, 
its stationary memoryless extension $\chn$ is a channel 
which maps an $n$-tuple $(\sigma_1, \ldots , \sigma_n)
\in\cXn=\cXtn$ 
of states $\{\sigma_i\}\subset\cS({\cH}_1)$ 
to the product state 
$\Gamma(\sigma_1)\otimes\cdots\otimes\Gamma(\sigma_n)$, 
and the capacity of $\chvec =\{\chn\}$ is given by 
(\ref{capacity_memoryless.1}). 
On the other hand, $\Gamma$ has the stationary 
memoryless extension $\Gamma^{\otimes n}$ as 
a ``quantum-quantum" channel, which defines another 
classical-quantum channel 
$\tilde{\ch}^{(n)} : \tilde{\cX}^{(n)}\rightarrow\cS ({{\cH}_2}^{\otimes n})$ 
with $\tilde{\cX}^{(n)} = \cS({{\cH}_1}^{\otimes n})$.  Note that 
$\chn$ can be regarded as the restriction $\tilde{\ch}^{(n)}\restrict{\cXn}$ of 
$\tilde{\ch}^{(n)}$ by identifying $(\sigma_1, \ldots , \sigma_n)\in\cXtn$ 
with $\sigma_1\otimes\cdots\otimes\sigma_n\in\tilde{\cX}^{(n)}$. 
The capacity $C(\vec{\tilde{\bm{\ch}}})$ of 
$\vec{\tilde{\bm{\ch}}}=\{\tilde{\ch}^{(n)}\}$ 
is beyond the scope of the preceding theorem, whereas 
recently the conjecture 
$C(\vec{\tilde{\bm{\ch}}})=C(\chvec)$ together with the 
more fundamental additivity conjecture has been 
calling wide attention. See, for instance, \cite{OsaNag, Shor, King1, King2} 
and the references cited there. 
\end{remark}

Historically, the converse part $C (\chvec) \leq 
\sup_{P\in\cP (\cX)} I(P,\ch)$ 
was first established by Holevo's early work 
\cite{Holevo-bounds, Holevo-bounds2} 
which is now often referred to as the Holevo bound, 
while the direct part $C(\chvec) \geq \sup_{P\in\cP
(\cX)} I(P,\ch)$ was proved much more recently by Holevo \cite{Holevo-QCTh} 
and Schumacher-Westmoreland \cite{Schumacher-Westmoreland}. 
It should be noted that their proof 
is based on the representation of $I(P,\ch)$ as the 
entropy difference: 
\begin{equation}
I(P,\ch) = H(\charg{P}) - \sum_{x} P(x) H(\charg{x}),
\end{equation} 
where $H(\rho)\defeq -\Tr[\rho\log\rho]$ is the von Neumann entropy, 
and hence needs (when $\dim\cH =\infty$) the assumption 
\begin{equation}
\label{eq:finiteH}
H(\charg{x}) < \infty, \quad \forall x\in\cX.
\end{equation}
See the next section for more details.  Our proof given below 
has the advantage of not needing this finiteness assumption
(cf.\ Remark~\ref{remark:finiteH}). 
Note also that in the case when 
$\dim\cH<\infty$ 
the range of 
supremum in (\ref{capacity_memoryless.1}) 
can be restricted to those $P\in\cP(\cX)$ with 
$\left|\,{\rm supp} (P)\right|\leq \dim\Delta +1$, where 
$\left|\,{\rm supp} (P)\right|$ denotes the number of elements 
of the support of $P$ and $\Delta\defeq\{\charg{x}\,|\,x\in\cX\}$, 
and that the supremum can be replaced with maximum when $\Delta$ 
is closed (and hence compact); see \cite{Fuj-Nag, Uhlmann}. 
The strong converse $C^\dagger (\chvec) \leq 
\sup_{P\in\cP (\cX)} I(P,\ch)$ for a finite $\cX$
was shown in \cite{Oga-Nag:channel, Winter}. 

Let us begin with considering the (weak) converse 
\begin{equation}
\label{weak_conv_memoryless}
C(\chvec) \leq \sup_{P\in\cP(\cX)} I(P,\ch).
\end{equation}

\begin{lemma} 
\label{lemma:infI=<liminf}
For any sequence of channels 
$\chvec=\{\chn\}$ and any sequence of distributions 
$\Pvec=\{\Pn\}$ we have
\begin{align}
\infI (\Pvec, \chvec) &\leq
\liminf_{n\rightarrow\infty}
\frac{1}{n} I(\Pn,\chn).
\end{align}
\end{lemma}

\begin{proof} 
Given $n$, $\xn\in\cXn$ and $a\in\bR$ arbitrarily, 
let
\begin{align*}
\alpha_n &\defeq \Tr\left[\chnarg{\xn}
\spec{\chnarg{\xn}-e^{na}\chnarg{\Pn} >0}\right], \\
\beta_n &\defeq \Tr\left[\chnarg{\Pn}
\spec{\chnarg{\xn}-e^{na}\chnarg{\Pn} >0}\right].
\end{align*}
Then the monotonicity of the quantum relative entropy 
yields
\begin{align*}
D(\chnarg{\xn}\,\|\,\chnarg{\Pn}) &\geq
\alpha_n \log\frac{\alpha_n}{\beta_n} + 
(1-\alpha_n)\log\frac{1-\alpha_n}{1-\beta_n} \\
&\geq
-\log 2 -\alpha_n\log \beta_n.
\end{align*}
On the other hand, we have
\[
0\leq \Tr\left[
(\chnarg{\xn} -e^{na}\chnarg{\Pn} )
\spec{\chnarg{\xn} -e^{na}\chnarg{\Pn} >0}
\right] 
= \alpha_n - e^{na} \beta_n
\]
and hence $\beta_n\leq e^{-na}\alpha_n\leq e^{-na}$. 
We thus obtain 
$
\frac{1}{n} D(\chnarg{\xn}\,\|\,\chnarg{\Pn})\geq
-\frac{1}{n}\log 2 + a \,\alpha_n
$, 
and taking the expectation w.r.t.\ $\Pn$ we have
\[
\frac{1}{n} I(\Pn,\chn) 
\geq
-\frac{1}{n}\log 2 + 
a \sum_{\xn} \Pn(\xn) 
\Tr\left[\chnarg{\xn}
\spec{\chnarg{\xn}-e^{na}\chnarg{\Pn} >0}\right].
\]
This leads to the implications:
\begin{align*}
a <\infI(\Pvec\,\|\,\chvec) 
&\Longrightarrow\;
\lim_{n\rightarrow\infty} 
 \sum_{\xn} \Pn(\xn) \Tr\left[\chnarg{\xn}
\spec{\chnarg{\xn}-e^{na}\chnarg{\Pn} >0}\right]
=1 \\
&\Longrightarrow\;
a\leq \liminf_{n\rightarrow\infty} 
\frac{1}{n} I(\Pn,\chn), 
\end{align*}
which proves the lemma.
\end{proof}

Using this lemma and invoking that 
in the stationary memoryless case
\[
\sup_{\Pn\in\cP ({\cX}^n)} I (\Pn, \chn) 
= 
n \sup_{P\in\cP(\cX)} I(P, \ch),
\]
we see that (\ref{weak_conv_memoryless}) follows 
from the general formula 
$C(\chvec)\leq\max_{\Pvec} \infI (\Pvec,\chvec)$. 

Before proceeding to the direct and strong converse parts, 
we introduce quantum analogues 
of the spectral inf- and sup-divergence rates \cite{Han_book} 
(see \ Remark~\ref{remark:J>=I}): 
given arbitrary sequences of states $\rhovec=\{\rhon\}$ 
and $\sigmavec=\{\sigman\}$, let 
\begin{align}
\supD (\rhovec\,\|\,\sigmavec) 
&\defeq 
\inf\,\bigl\{a\,|\, \lim_{n\rightarrow\infty} 
\Tr \left[ \rhon \spec{\rhon - e^{na}\sigman >0}
\right] = 0\bigr\} ,\\
\infD (\rhovec\,\|\,\sigmavec) 
&\defeq 
\sup\,\bigl\{a\,|\, \lim_{n\rightarrow\infty} 
\Tr \left[ \rhon \spec{\rhon - e^{na}\sigman >0}
\right] = 1\bigr\}.
\end{align}
Note that $\infD (\rhovec\,\|\,\sigmavec) \leq 
\supD (\rhovec\,\|\,\sigmavec)$ and that 
$\infD (\rhovec\,\|\,\sigmavec) \leq 
\liminf_{n\rightarrow\infty}
\frac{1}{n} D(\rhon\,\|\,\sigman)$, the latter 
of which can be proved similarly to Lemma~\ref{lemma:infI=<liminf}. 
The following relation, which was shown in \cite{Nag-Hay:test},
 will play an essential role in the later arguments:
in the quantum i.i.d.\ case when 
$\rhotvec = \{\rhotensor\}_{n=1}^\infty$ and 
$\sigmatvec=\{\sigmatensor\}_{n=1}^\infty$, 
we have 
\begin{align}
\Label{divergence_rate_q_iid}
\supD(\rhotvec\,\|\,\sigmatvec) = 
\infD(\rhotvec\,\|\,\sigmatvec) =
D(\rho\,\|\,\sigma) .
\end{align}

Now let us observe how the direct part
\begin{align}
\Label{direct_memoryless}
C(\chvec) \geq 
\sup_{P\in\cP (\cX)} I(P,\ch) 
\end{align}
follows from the general formula. 
Let $P$ be an arbitrary distribution in $\cP (\cX)$ 
and $\Ptn \in \cP (\cXtn)$ be the 
$n$th i.i.d.\ extension:
$\Ptn (\xn) = P(x_1)\cdots P(x_n)$ 
for $\xn=(x_1, \ldots , x_n)$. 
Denoting the support of $P$ by 
$\{u_1, \ldots , u_k\}\subset\cX$ and letting 
$\lambda_i =P(u_i)$, $\rho_i=\charg{u_i}$ and 
$\sigma=\charg{P}= \sum_i\lambda_i\rho_i$, 
 we have
\begin{align*}
&\sum_{\xn\in\cXn} 
\Ptn(\xn) \Tr \left[ \chnarg{\xn} 
\spec{\chnarg{\xn} - e^{na}\chnarg{\Ptn} >0}
\right] \\
&=
\sum_{i_1, \ldots , i_n} \lambda_{i_1}\cdots \lambda_{i_n}
\Tr \left[ (\rho_{i_1} \otimes\cdots\otimes \rho_{i_n} )
\spec{\rho_{i_1}\otimes\cdots\otimes \rho_{i_n} 
- e^{na} \sigma^{\otimes n} >0} \right] \\
&= \Tr \left[ R^{\otimes n} 
\spec{ R^{\otimes n} - e^{na} 
S^{\otimes n} > 0} \right], 
\end{align*}
where
\begin{align}
\label{def:R_S}
R\defeq 
\left( \begin{array}{ccc}
\lambda_1 \rho_1 & &  \smash{\lower1.4ex\hbox{0}}
\\
\smash{\lower1.7ex\hbox{0}} & \ddots & \\
&& \lambda_k \rho_k
\end{array}
\right), \qquad 
S \defeq 
\left( \begin{array}{ccc}
\lambda_1 \sigma & &  \smash{\lower1.4ex\hbox{0}}
\\
\smash{\lower1.7ex\hbox{0}} & \ddots & \\
&& \lambda_k \sigma
\end{array}
\right) .
\end{align}
We thus have for the sequences $\Ptvec=\{\Ptn\}, 
\vec{\bm{R}} =\{R^{\otimes n}\}$ and 
$\vec{\bm{S}} =\{S^{\otimes n}\}$
\begin{align}
\Label{infI=I}
\infI (\Ptvec, \chvec) = 
\infD (\vec{\bm{R}}\,\|\,\vec{\bm{S}}) 
= D(R\,\|\,S) =
I(P,\ch) ,
\end{align}
where 
the second equality follows from (\ref{divergence_rate_q_iid}) 
and the rest are immediate from the definitions 
of the quantities. 
This, combined with (\ref{generalC.1}),  
completes the proof of (\ref{direct_memoryless}). 

\begin{remark}
\label{remark:OD2}
\rm
Essential in the above derivation of (\ref{direct_memoryless}) 
from (\ref{generalC.1}) is the use of  
$\infD(\rhovec\,\|\,\sigmavec)\geq
D(\rho\,\|\,\sigma)$ for sequences of i.i.d.\ states. 
The proof of the inequality given in \cite{Nag-Hay:test} 
is based on the direct part 
of the quantum Stein's lemma
for a hypothesis testing problem 
on $\rhotensor$ and $\sigmatensor$, 
which was first shown 
by Hiai and Petz \cite{Hiai-Petz}, 
whereas the classical counterpart of the inequality is 
a direct consequence of the weak law of large numbers. 
Hence the above derivation can be thought of as a proof 
of the channel coding theorem via the theory of 
quantum hypothesis testing 
(cf.\ Remark~\ref{remark:ch_from_hypo} below).
It should be noted that 
a significant characteristic of the proof lies in 
separation of the coding part and 
the limiting part; the former is entirely coped with 
in the general formula (\ref{generalC.1}), or equivalently  
in the non-asymptotic arguments of 
Lemma~\ref{lemma:ineq_for_direct} and 
Lemma~\ref{lemma:direct}, while  
the latter relies on the asymptotic analysis of quantum hypothesis testing. 
Another proof of  (\ref{direct_memoryless})  
with a similar approach is found in \cite{OD, ON1}, where 
the coding part is proved by a variant of quantum 
Feinstein's lemma (cf.\ Remark~\ref{remark:OD1}) 
and the limiting part is based on 
an asymptotic analysis made in \cite{OH1} 
(cf.\ Remark~\ref{remark:ch_from_OH1} below) 
on a variant of 
$\infD(\rhovec\,\|\,\sigmavec)$, 
which is much easier to treat than the original 
$\infD(\rhovec\,\|\,\sigmavec)$. 
\end{remark}

\begin{remark}
\label{remark:ch_from_hypo}
\rm In an actual fact,  (\ref{direct_memoryless}) 
can be proved by directly applying Lemma~\ref{lemma:ineq_for_direct} to 
the direct part of quantum Stein's lemma as follows, without 
appealing to the general formula (\ref{generalC.1}). 
Given $P\in\cP(\cX)$, 
let $R$ and $S$ be defined by (\ref{def:R_S}), which can be 
represented as 
$R=\oplus_{x}P(x)\charg{x}$ and 
$S=\oplus_{x} P(x) \charg{P}$. 
For an arbitrary $\varepsilon>0$ and a sufficiently large $n$, 
it follows from the quantum Stein's lemma that 
there exists a projection of the form $T^{(n)}=
\oplus_{\xn}T^{(n)}_{\xn}$, where $\{T^{(n)}_{\xn}\}$ are projections 
on $\cHtensor$, such that 
\begin{align*}
\Tr [R^{\otimes n} T^{(n)}] 
&= \sum_{\xn} \Pn(\xn) 
\Tr[\chnarg{\xn} T^{(n)}_{\xn}] 
\geq 1-\varepsilon, \\
\Tr [S^{\otimes n} T^{(n)}] 
&= \sum_{\xn} \Pn(\xn) 
\Tr[\charg{P}^{\otimes n} T^{(n)}_{\xn}] 
\leq 
e^{-n(D(R\,\|\,S)-\varepsilon)} .
\end{align*}
Given an encoder $\encn :\{1, \ldots , N\}\rightarrow\cXtn$, 
define the decoding POVM $Z^{(n)}=\{Z^{(n)}_i\}_{i=1}^N$ by 
\[
Z_i^{(n)}\defeq
\left(
\sum_{j=1}^N T^{(n)}_{\encnarg{j}}
\right)^{-\frac{1}{2}} 
T^{(n)}_{\encnarg{i}}
\left(
\sum_{j=1}^N T^{(n)}_{\encnarg{j}}
\right)^{-\frac{1}{2}}.
\]
Then replacing $\decn$ with $Z^{(n)}$ in the proof of 
Lemma~\ref{lemma:direct}, using 
Lemma~\ref{lemma:ineq_for_direct} for $c=1$ (e.g.) 
 and applying the random coding with respect to $\Pn$, 
we see that there exists a code $\coden$ satisfying 
\begin{align}
\Pe[\coden]
&\leq
2 \left(1-\Tr[R^{\otimes n} T^{(n)}]\right) 
+
4 N \Tr[ S^{\otimes n} T^{(n)}] 
\label{ineq:direct_ch_hypo}
\\
&\leq
2 \varepsilon + 4 e^{-n (D(R\,\|\,S)-\varepsilon)} N ,
\nonumber
\end{align}
which proves (\ref{direct_memoryless}) by 
$D(R\,\|\,S)=I(P,\ch)$. 
\end{remark}

\begin{remark}
\label{remark:finiteH}
\rm 
As is shown in section 4 of \cite{Hiai-Petz}, 
from the fact that the (direct part of) quantum 
Stein's lemma holds for states on every finite-dimensional 
matrix algebra, it is immediately concluded that the lemma 
holds also for states on every 
AFD (approximately finite dimensional)  operator algebra, 
including the algebra $B(\cH)$ of bounded operators on 
a separable Hilbert space $\cH$.  
This means that 
our proof of  (\ref{capacity_memoryless.1}) 
is valid for every channel $\ch$ on a separable Hilbert space 
$\cH$ without the finiteness assumption (\ref{eq:finiteH}). 
Note also that a similar argument based on the AFD property can be applied to 
the channel coding problem directly to remove the finiteness assumption 
from the proof of Holevo-Schumacher-Westmoreland. 
\end{remark}

\begin{remark}
\label{remark:ch_from_OH1}
\rm 
Combination of the argument in Remark~\ref{remark:ch_from_hypo} 
and the derivation of the direct part of 
quantum Stein's lemma given in \cite{OH1} 
will provide one of the simplest proofs of (\ref{direct_memoryless})
(for a finite-dimensional $\cH$). 
In addition, application of Theorem~2 of \cite{OH1} to 
(\ref{ineq:direct_ch_hypo}) implies that for any $n$ and $a>0$ there 
exists a code $\coden$ satisfying $|\coden|= e^{na}$ and 
\begin{equation}
\Pe[\coden]
\leq
6 (n+1)^d e^{-n \, \bar{\varphi}(a)},
\label{ineq_Pe_OH1}
\end{equation}
where $d\defeq k\dim\cH$ (the size of the matrices $R$ and $S$) 
and 
\begin{align*}
\bar{\varphi} (a) &\defeq \max_{0\leq t\leq 1} 
\left(-a t-\log \Tr\left[R S^{\frac{t}{2}} R^{-t} S^{\frac{t}{2}}
\right]\right) \\
&=
\max_{0\leq t\leq 1} 
\left(-a t-\log \sum_i \lambda_i 
\Tr\left[\rho_i \sigma^{\frac{t}{2}} \rho_i^{-t} \sigma^{\frac{t}{2}}
\right]\right).
\end{align*}
As was shown in \cite{OH1}, $\bar{\varphi}(a) >0$ 
holds for any $a < D(R\,\|\,S) =I(P,\ch)$, and 
(\ref{ineq_Pe_OH1}) gives an exponential bound 
on the error probability. 
\end{remark}

Next we proceed to the strong converse part 
\begin{align}
\Label{converse_memoryless}
C^\dagger(\chvec) 
\leq 
\sup_{P\in\cP (\cX)} I(P,\ch) 
\end{align} 
under the assumption that $\cH$ is finite-dimensional. 
In order to link 
(\ref{converse_memoryless}) to the general formula, 
we use the following relations 
(\cite{Ohy-Pet-Wat, Sch-Wes:optimal}):
\begin{align} 
\nonumber
\sup_{P\in\cP (\cX)} I(P,\ch) &= 
\sup_{P\in\cP (\cX)} \min_{\sigma\in\cS(\cH)} 
J(P,\sigma,\ch) \\
\nonumber
&= \min_{\sigma\in\cS(\cH)} \sup_{P\in\cP (\cX)}
J(P,\sigma,\ch) \\
&=\min_{\sigma\in\cS (\cH)} \sup_{x\in\cX}\, 
D(\charg{x}\,\|\,\sigma) ,
\Label{capacity_memoryless.2}
\end{align}
where 
\[ 
J(P,\sigma ,\ch)\defeq \sum_{x\in\cX} P(x) D(\charg{x}\,\|\,\sigma).
\]
These relations can be derived just in parallel with its 
classical counter part (e.g., pp.142--147 of \ \cite{Csi-Kor}, 
 Theorem 4.5.1 of \cite{Gal}) 
by the use of a mini-max theorem for a certain class of 
two-variable convex-concave 
functions (e.g.\ Chap.VI of \cite{Eke-Tem}), 
combined with the fact that the supremum of 
$\sup_{P\in\cP(\cX)} I(P, \ch)$ can be attained when 
$\{\charg{x}\,|\,x\in\cX\}$ is closed (\cite{Fuj-Nag, Uhlmann}). 

In proving the strong converse of the quantum 
hypothesis testing problem for two i.i.d.\ states, which 
is equivalent to the part $\supD (\rhotvec\,\|\,\sigmatvec) 
\leq D(\rho\,\|\,\sigma)$ in (\ref{divergence_rate_q_iid}) 
(see \cite{Nag-Hay:test}), Ogawa and Nagaoka \cite{Oga-Nag:test}
showed that for any states $\rho, \sigma$ and 
any numbers $c>0$ and $0\leq s\leq 1$, 
\begin{equation}
\label{ineq:OgaNag}
\Tr \left[ \rho \spec{\rho - c \sigma >0}\right] 
\leq c^{-s} \Tr \left[ \rho^{1+s} \sigma^{-s}\right] . 
\end{equation}
Applying this to the states $\chnarg{\xn}$, $\sigmatensor$
and 
$c=e^{na}$, we have
\begin{align}
\Tr \left[ \chnarg{\xn} 
\spec{ \chnarg{\xn} - e^{na} \sigmatensor >0} \right] 
&\leq \exp\Bigl[ -n \Bigl(as - \frac{1}{n} \sum_{i=1}^n
 \log \Tr \left[ \charg{x_i}^{1+s} \sigma^{-s}\right] \Bigr)
\Bigr] 
\nonumber 
\\
&\leq \exp\Bigl[ -n \Bigl(as - \sup_{x\in\cX}
 \log \Tr \left[ \charg{x}^{1+s} \sigma^{-s}\right] \Bigr)
\Bigr] .
\Label{ineq_for_strongconv}
\end{align}
Now assume that
$\mbox{Im}\,\sigma \supset \mbox{Im}\,\charg{x}$ for 
all $x\in\cX$, where $\mbox{Im}$ denotes the image 
(range) of an operator, 
let $\bar{\Delta}$ be the closure of the 
range $\Delta=\{\charg{x}\,|\,x\in\cX\}$, 
and define the function $f :[0,1] \times \bar{\Delta}
\rightarrow \bR$ by 
$f(s, \rho) = \log  \Tr \left[ \rho^{1+s} \sigma^{-s}\right] $.
Then we have  $f(0, \rho) =0$ and
\begin{align}
\Label{diff_f_D}
\frac{\partial}{\partial s} f(0, \rho) = D(\rho\,\|\,\sigma) .
\end{align}
Moreover, since the derivative 
\[
\frac{\partial}{\partial s} f(s, \rho) = 
\frac{\Tr \left[ \rho^{1+s}(\log\rho - \log\sigma) \sigma^{-s}\right]}
{\Tr \left[ \rho^{1+s}\sigma^{-s}\right]}
\]
is continuous with respect to both $s$ and $\rho$, and since 
$\bar{\Delta}$ is compact, we see that the differentiation in 
(\ref{diff_f_D}) is uniform in $\rho$; i.e., 
\begin{align*}
\lim_{s\downarrow 0} 
\max_{\rho\in\bar{\Delta}} \left|\frac{f(s,\rho)}{s} - D(\rho\,\|\,\sigma)\right| 
=0 .
\end{align*}
Let $a$ be an arbitrary number satisfying 
$a > \max_{\rho\in\bar{\Delta}} D(\rho\,\|\,\sigma) 
=\sup_{x\in\cX} D(\charg{x}\,\|\,\sigma)$. 
It then follows from the above uniform convergence that there exists an $s_0 >0$ such 
that for any $0 < s \leq s_0$
\[ as > \max_{\rho\in\bar{\Delta}} f(s,\rho) = 
\sup_{x\in\cX} \log  \Tr \left[ \charg{x}^{1+s} \sigma^{-s}\right] .
\]
Invoking (\ref{ineq_for_strongconv}), this implies that for 
any sequence $\xvec =\{\xn\}\in\cXvec$, where 
$\cXvec=\{\cXtn\}$ is identified with the product set 
$\prod_n\cXtn$, we have
\begin{equation}
\Label{equivalent:supD=<maxD}
\lim_{n\rightarrow\infty} \Tr \left[ \chnarg{\xn} 
\spec{ \chnarg{\xn}  - e^{na} \sigmatensor >0} \right]  
=0\quad\mbox{for}\;\;
\forall a >  \sup_{x\in\cX} D(\charg{x}\,\|\,\sigma), 
\end{equation}
or equivalently
\begin{align}
\Label{supD=<maxD}
\supD (\chvecarg{\xvec}\,\|\,\sigmatvec) \leq 
\sup_{x\in\cX} D(\charg{x}\,\|\,\sigma) ,
\end{align}
where $\chvecarg{\xvec}=\{\chnarg{\xn}\}$ 
and $\sigmavec=\{\sigmatensor\}$.  
Although we assumed  
$\mbox{Im}\,\sigma \supset \mbox{Im}\,\charg{x}$, $\forall x\in\cX$ 
above, this inequality is valid for any $\sigma\in\cS (\cH)$ 
because 
$\sup_{x\in\cX} D(\charg{x}\,\|\,\sigma) = \infty$ if 
$\mbox{Im}\,\sigma \not\supset \mbox{Im}\,\charg{x}$ for some $x\in\cX$. 
Now the desired inequality (\ref{converse_memoryless}) is 
derived from the general formula (\ref{generalC+.2}) as follows:
\begin{align*}
C^\dagger(\chvec) 
&= 
\max_{\Pvec\in\cPvec(\cXvec)} \; \min_{\sigmavec\in\cSvec(\cHvec)} \; 
\supJ (\Pvec, \sigmavec, \chvec)
\\ &\leq
\min_{\sigmavec\in\cSvec(\cHvec)} \; 
\max_{\Pvec\in\cPvec(\cXvec)} \; 
\supJ (\Pvec, \sigmavec, \chvec)
\\ &=
\min_{\sigmavec\in\cSvec(\cHvec)} 
\max_{\xvec\in\cXvec} \;
\supD (\chvecarg{\xvec}\,\|\,\sigmavec)  
\\ &\leq 
\min_{\sigma\in\cS (\cH)}
\max_{\xvec\in\cXvec} \;
\supD (\chvecarg{\xvec}\,\|\,\sigmatvec)  
\quad\mbox{with}\quad
\sigmavec =\{\sigmatensor\}
\\ &\leq
\min_{\sigma\in\cS (\cH)}
\,\sup_{x\in\cX} D(\charg{x}\,\|\,\sigma) 
=
\sup_{P\in\cP (\cX)} I(P,\ch) ,
\end{align*}
where the last equality follows from (\ref{capacity_memoryless.2}).

\section{On the Holevo-Schumacher-Westmoreland decoder}
\Label{s7}

Let us return to the situation in the proof of Lemma \ref{lemma:direct} 
where a probability distribution $\Pn$ and an encoder 
$\encn:\{1, \ldots , N\}\rightarrow\cXn$ are given.  
Instead of $\decn$ defined in (\ref{def_Xni}), 
consider the following POVM $\decvarn$:
\begin{equation}
\decvarnarg{i} \defeq \left(\sum_{j=1}^N 
\tau \nu_j \tau
\right)^{-\frac{1}{2}}
\tau \nu_i \tau
\left(\sum_{j=1} ^N 
\tau \nu_j \tau \right)^{-\frac{1}{2}} ,
\label{def_tildY}
\end{equation}
where
\[
\tau \defeq 
\spec{\chnarg{\Pn} < e^{-n b}}, \quad
\nu_i \defeq 
\spec{\chnarg{\encnarg{i}} > e^{-nc}} .
\]
This type of decoder was introduced by Holevo \cite{Holevo-QCTh} 
and Schumacher-Westmoreland \cite{Schumacher-Westmoreland} 
in proving the direct part of the capacity theorem. 
Let us investigate this decoder, comparing it with 
our $\decnarg{i}$ defined by (\ref{def_Xni}) and (\ref{def_pii}). 

\begin{remark}
\rm 
More precisely, the decoder treated in \cite{Holevo-QCTh, Schumacher-Westmoreland} 
was defined by (\ref{def_tildY}) with 
projections $\tau$ and $\nu_i$ of the form
\[
\tau \defeq 
\spec{e^{-n b'}< \chnarg{\Pn} < e^{-n b}}, \quad
\nu_i \defeq 
\spec{e^{-nc}< \chnarg{\encnarg{i}} <e^{-nc'}} ,
\]
where we have used a slight 
extension of  the notation in (\ref{12-13-1}): 
\[
\spec{\alpha <A<\beta}= \sum_{i:\, \alpha<\lambda_i<\beta} E_i.
\]
However, the asymptotic performance of the decoder does not 
depend on the particular values of $b', c'$ as far as 
$b'$ is sufficiently large and $c'$ is sufficiently small.  
Hence we set $b'=\infty$ and $c'=-\infty$ to simplify the 
arguments. 
\end{remark}

The authors of \cite{Holevo-QCTh, Schumacher-Westmoreland} 
showed by a rather complicated
calculation that  the average error probability of the code $\codevarn =(N, \encn, \decvarn)$ 
satisfies
\begin{align}
\label{ineq1:HolSchWes}
\Pe[\codevarn]
\leq 
\frac{1}{N} \sum_{i=1}^N \left\{
3 \Tr\left[\chnarg{\encnarg{i}} (I-\tau) \right] + 
\Tr\left[\chnarg{\encnarg{i}} (I-\nu_i)\right]
+ 
\sum_{j(\neq i)} \Tr\left[\chnarg{\encnarg{i}}\, \tau\nu_j\tau\right]
\right\} .
\end{align}
Note that a simplified derivation of the inequality 
with slightly different coefficients was shown 
in \cite{Hol:quant-ph9809023}.  
Applying the random coding with respect to $\Pn$ to 
(\ref{ineq1:HolSchWes}) and noting that 
\begin{align}
\nonumber
&\Tr\left[\chnarg{\Pn}
\spec{\chnarg{\Pn}<e^{-nb}}\spec{\chnarg{\xn}>e^{-nc}}\spec{\chnarg{\Pn}<e^{-nb}}
\right] \\
\nonumber
&\leq 
\, \left\|\chnarg{\Pn}\spec{\chnarg{\Pn}<e^{-nb}}\right\| 
\cdot
 \Tr\left[\spec{\chnarg{\xn}>e^{-nc}}\right] \\
&\leq 
e^{-n(b-c)},
\label{ineq_for_HolSchWes}
\end{align}
where $\|\cdot\|$ denotes the operator norm, 
we see that there exists a code $\coden$ such
that
\begin{align}
\nonumber
\Pe [\coden] \leq&
3 \Tr\left[\chnarg{\Pn}
\spec{\chnarg{\Pn}\geq e^{-nb}}\right] + 
\sum_{\xn} \Pn(\xn) \Tr\left[ 
\chnarg{\xn}\spec{\chnarg{\xn}\leq e^{-nc}}\right]
\\
&+ 
e^{-n(b-c)} N .
\label{ineq:direct_HolSchWes}
\end{align}
Now, for an arbitrary $\Pvec=\{\Pn\}\in\cPvec(\cXvec)$ let 
\begin{align*}
 \infH (\chvecarg{\Pvec})
&\defeq \sup\,\left\{b\,\left|\, 
\lim_{n\rightarrow\infty} \Tr\left[\chnarg{\Pn}
\spec{\chnarg{\Pn}\geq e^{-nb} }\right]=0 \right.\right\} \\
&= \sup\,\left\{b\,\left|\, 
\lim_{n\rightarrow\infty} \Tr\left[\chnarg{\Pn}
\spec{-\frac{1}{n}\log \chnarg{\Pn}\leq b }\right]=0 \right.\right\},
\end{align*}
\begin{align*}
\supH (\chvec\,|\,\Pvec) 
&\defeq \inf\,\left\{c\,\left|\, 
\lim_{n\rightarrow\infty} 
\sum_{\xn} \Pn (\xn) 
\Tr\left[\chnarg{\xn}
\spec{\chnarg{\xn} \leq e^{-nc} }\right]=0 \right.\right\} \\
&= \inf\,\left\{c\,\left|\, 
\lim_{n\rightarrow\infty} 
\sum_{\xn} \Pn (\xn) 
\Tr\left[\chnarg{\xn}
\spec{-\frac{1}{n}\log\chnarg{\xn} \geq c }\right]=0 \right.\right\},
\end{align*}
and assume that $\supH (\chvec\,|\,\Pvec) <\infty$.  
It then follows from (\ref{ineq:direct_HolSchWes}) that 
there exists a sequence of codes $\codevec =\{\coden\}$ 
such that $\lim_{n\rightarrow\infty} \Pe [\coden] =0$ 
with the rate $\liminf_{n\rightarrow\infty}\frac{1}{n}
\log |\coden |$ being arbitrarily close to 
$\infH (\chvecarg{\Pvec}) - \supH (\chvec\,|\,\Pvec)$; 
i.e., we have
\begin{equation}
\label{ineq2:HolSchWes}
 C (\chvec) \geq 
\max_{\Pvec\in\cPvec(\cXvec)} \; 
\left\{
\left.
\infH (\chvecarg{\Pvec}) - 
\supH (\chvec\,|\,\Pvec) 
\;\right|\; \supH (\chvec\,|\,\Pvec) < \infty
\right\}.
\end{equation}

The quantities $\infH (\chvecarg{\Pvec})$ and 
$\supH (\chvec\,|\,\Pvec)$ are regarded as 
information-spectrum analogues of the von Neumann entropy and 
its conditional version.  Indeed, for 
a stationary memoryless channel 
$\chnarg{\xn}=\charg{x_1}\otimes\cdots\otimes\charg{x_n}$ 
with i.i.d. $\Pn (\xn)=
P(x_1)\cdots P(x_n)$ the law of 
large numbers yields 
\begin{align} 
\infH (\chvecarg{\Pvec}) &= 
H(\charg{P})=-\Tr[\charg{P}\log\charg{P}], 
\label{eq:infH=H}
\\
\supH (\chvec\,|\,\Pvec) &= 
H(\ch\,|\,P) \defeq
\sum_{x} P(x) H(\charg{x}),
\label{eq:supH=H}
\end{align}
which leads to $C(\chvec) \geq 
\sup_{P\in\cP(\cX)}\left(
H(\charg{P}) - H(\ch\,|\,P) \right)
=
\sup_{P\in\cP(\cX)} I(P,\ch)$ 
under the finiteness assumption (\ref{eq:finiteH})  
(cf.\ Remark~\ref{remark:finiteH}). 
This is just what was shown in \cite{Holevo-QCTh, Schumacher-Westmoreland}. 

\begin{remark}
\label{remark:simplify_HSW}
{\em 
Inequality (\ref{h1}) of Lemma~\ref{lemma:ineq_for_direct} 
can be applied to the code $\codevarn =(N, \encn, \decvarn)$ 
to derive (\ref{ineq2:HolSchWes}) 
more straightforwardly 
than the  derivations in 
\cite{Holevo-QCTh, Schumacher-Westmoreland, Hol:quant-ph9809023}. 
Indeed, letting $c=1$ (e.g.)\ in (\ref{h1}) 
we have
\begin{align*}
\Pe[\codevarn] 
&
\leq 
\frac{1}{N} \sum_{i=1}^N \left\{
2 \Tr\left[\chnarg{\encnarg{i}}\, 
(I-\tau\nu_i\tau) \right]
+
4 \sum_{j\neq i}\Tr\left[\chnarg{\encnarg{i}}\,\tau\nu_j\tau\right]
\right\} \\
&
\leq 
\frac{1}{N} \sum_{i=1}^N \left\{
 4 \Tr\left[\chnarg{\encnarg{i}} (I-\tau) \right] + 
2 \Tr\left[\chnarg{\encnarg{i}} (I-\nu_i)\right]
+ 
4 \sum_{j(\neq i)} \Tr\left[\chnarg{\encnarg{i}}\, \tau\nu_j\tau\right]
\right\} ,
\end{align*}
where the second inequality follows from 
the next lemma.  This leads to (\ref{ineq2:HolSchWes}) 
as well as from (\ref{ineq1:HolSchWes}). 
}
\end{remark}

\begin{lemma}
\label{lemma:rhotaunutau}
For any state $\rho$ and any projections $\nu, \tau$ such that 
$[\rho, \nu]=0$, we have
\[
\Tr[\rho \tau\nu\tau] 
\geq
\Tr[\rho\nu] -2 \Tr[\rho (I-\tau)].
\] 
\end{lemma}

\begin{proof} 
Obvious from 
$
0\leq (I-\tau)\nu(I-\tau) = 
\tau\nu\tau -\nu + (I-\tau)\nu + \nu (I-\tau)$
and
$\rho\nu = \nu\rho \leq \rho$. 
\end{proof}

Comparing (\ref{ineq2:HolSchWes}) with  
(\ref{generalC.1}) it is immediate that
\[
\max_{\Pvec}\infI(\Pvec,\chvec) 
\geq 
\max_{\Pvec} \; 
\left\{
\left.
\infH (\chvecarg{\Pvec}) - 
\supH (\chvec\,|\,\Pvec) 
\;\right|\; \supH (\chvec\,|\,\Pvec) < \infty
\right\}. 
\]
Actually, a slightly stronger 
assertion holds:

\begin{theorem}
\label{thm:infI>=infH-supH}
For every $\Pvec\in\cPvec(\cHvec )$ 
with $\supH (\chvec\,|\,\Pvec)<\infty$ we have
\begin{equation}
\label{eq:infI>=infH-supH}
\infI(\Pvec,\chvec) \geq 
\infH(\chvecarg{\Pvec}) - 
\supH (\chvec\,|\,\Pvec).
\end{equation}
\end{theorem}

\begin{proof}
It suffices to 
show that for any $b<\infH(\chvecarg{\Pvec})$, 
$c>\supH (\chvec\,|\,\Pvec)$ and $\varepsilon >0$ we have 
$\infI(\Pvec,\chvec) \geq b-c-\varepsilon$, or, equivalently that 
if 
\begin{equation}
\label{a1:infI>=infH-supH}
\lim_{n\to\infty}
\Tr\left[\chnarg{\Pn}\spec{\chnarg{\Pn}\geq e^{-nb}}\right]=0
\end{equation}
and
\begin{equation}
\label{a2:infI>=infH-supH}
\lim_{n\to\infty}
\sum_{\xn\in\cXn}\Pn(\xn) 
\Tr\left[\chnarg{\xn}\spec{\chnarg{\xn}> e^{-nc}}\right]=1
\end{equation}
then
\begin{equation}
\label{a3:infI>=infH-supH}
\lim_{n\to\infty}
\sum_{\xn\in\cXn}\Pn(\xn) 
\Tr\left[
\chnarg{\xn}\spec{\chnarg{\xn}- e^{n(b-c-\varepsilon)}\chnarg{\Pn}>0}\right] 
=1.
\end{equation}
We obtain
\begin{align*}
&\sum_{\xn\in\cXn}\Pn(\xn) 
\Tr\left[
\chnarg{\xn}\spec{\chnarg{\xn}- e^{n(b-c-\varepsilon)}\chnarg{\Pn}>0}\right] 
\\ \geq
&
\sum_{\xn\in\cXn}\Pn(\xn) 
\Tr\left[
\left(\chnarg{\xn}-e^{n(b-c-\varepsilon)}\chnarg{\Pn}\right)
\spec{\chnarg{\xn}- e^{n(b-c-\varepsilon)}\chnarg{\Pn}>0}\right] 
\\ \geq
&
\sum_{\xn\in\cXn}\Pn(\xn) 
\Tr\left[
\left(\chnarg{\xn}-e^{n(b-c-\varepsilon)}\chnarg{\Pn}\right)
\spec{\chnarg{\Pn}<e^{-nb}}
\spec{\chnarg{\xn}>e^{-nc}}
\spec{\chnarg{\Pn}<e^{-nb}}
\right] 
\\ \geq
&
\sum_{\xn\in\cXn}\Pn(\xn) 
\Tr\left[
\chnarg{\xn}
\spec{\chnarg{\xn}>e^{-nc}}
\right] 
-
2 
\sum_{\xn\in\cXn}\Pn(\xn) 
\Tr\left[\chnarg{\xn}\spec{\chnarg{\Pn}\geq e^{-nb}}\right] \\
&-
e^{n(b-c-\varepsilon)} 
\sum_{\xn\in\cXn}\Pn(\xn) 
\Tr\left[\chnarg{\Pn}\spec{\chnarg{\Pn}<e^{-nb}}\spec{\chnarg{\xn}>e^{-nc}}
\right] \\ \geq
&
\sum_{\xn\in\cXn}\Pn(\xn) 
\Tr\left[
\chnarg{\xn}
\spec{\chnarg{\xn}>e^{-nc}}
\right] 
-
2 
\Tr\left[\chnarg{\Pn}\spec{\chnarg{\Pn}\geq e^{-nb}}\right] 
- e^{-n\varepsilon} ,
\end{align*}
where the second inequality follows from 
(\ref{Neyman-Pearson}), 
the third from 
Lemma~\ref{lemma:rhotaunutau} 
and the last from (\ref{ineq_for_HolSchWes}). 
Now it is clear that (\ref{a2:infI>=infH-supH}) and 
 (\ref{a1:infI>=infH-supH}) implies 
(\ref{a3:infI>=infH-supH}). 
\end{proof}

\begin{remark}
\rm
Theorem~\ref{thm:infI>=infH-supH} 
enables us to derive the direct part (\ref{direct_memoryless}) 
for a stationary memoryless 
channel from 
the general formula (\ref{generalC.1}) 
via equations (\ref{eq:infH=H}) and (\ref{eq:supH=H}). 
This is essentially equivalent to 
the simplification of Holevo-Schumacher-Westmoreland's 
proof explained in 
Remark~\ref{remark:simplify_HSW}, but 
can also be regarded as a variation of the scenario of 
section~\ref{sec:memoryless} to derive (\ref{direct_memoryless}) 
from (\ref{generalC.1}) via $\infD(\rhovec\,\|\,\sigmavec)\geq
D(\rho\,\|\,\sigma)$ for $\rhotvec = \{\rhotensor\}_{n=1}^\infty$ and 
$\sigmatvec=\{\sigmatensor\}_{n=1}^\infty$.  
That is, 
just in parallel with the proof of Theorem~\ref{thm:infI>=infH-supH}, 
we can show for any sequences of states $\rhovec=\{\rhon\}$ and $\sigmavec=\{\sigman\}$ that 
\[
\infD (\rhovec\,\|\,\sigmavec) 
\geq 
\underline{K} (\rhovec\,\|\,\sigmavec) - 
\supH (\rhovec),
\] 
where 
\begin{align*}
\underline{K} (\rhovec\,\|\,\sigmavec) 
&\defeq 
\sup\,\left\{b\,\left|\, 
\lim_{n\rightarrow\infty} \Tr\left[\rhon
\spec{-\frac{1}{n}\log \sigman\leq b }\right]=0 \right.\right\},\\
\supH (\rhovec) 
&\defeq 
\inf\,\left\{c\,\left|\, 
\lim_{n\rightarrow\infty} 
\Tr\left[\rhon
\spec{-\frac{1}{n}\log\rhon \geq c }\right]=0 \right.\right\},
\end{align*}
which yields for 
$\rhotvec = \{\rhotensor\}_{n=1}^\infty$ and 
$\sigmatvec=\{\sigmatensor\}_{n=1}^\infty$ that 
$\infD(\rhovec\,\|\,\sigmavec)\geq 
- \Tr[\rho\log\sigma] - H(\rho) = 
D(\rho\,\|\,\sigma)$. 
Combination of this argument, which 
provides another simple proof of 
the direct part of the quantum Stein's lemma
(cf.\ \cite{Nag-Hay:test}), 
with the scenario of section~\ref{sec:memoryless} 
is equivalent to 
the direct use of Theorem~\ref{thm:infI>=infH-supH} mentioned above. 
\end{remark}

\begin{remark}
\rm 
The classical counterpart of (\ref{eq:infI>=infH-supH}) is 
rather obvious (cf.\ Remark~\ref{remark:J>=I}):
\begin{align*}
\infI (\bX\,;\, \bY) &= \mbox{p-}\liminf_{n\rightarrow\infty} 
\frac{1}{n} \log \frac{\Wn (\Yn\,|\,\Xn)}{P_{\Yn}(\Yn)} \\
&\geq 
\mbox{p-}\liminf_{n\rightarrow\infty} 
\frac{1}{n} \log \frac{1}{P_{\Yn}(\Yn) }
- 
\mbox{p-}\limsup_{n\rightarrow\infty} 
\frac{1}{n} \log \frac{1}{\Wn (\Yn\,|\,\Xn)} \\
&=
\infH (\bY) - \supH (\bY\,|\,\bX) .
\end{align*}
\end{remark}

\section{Capacity under cost constraint}\Label{s8}
\label{sec:cost}

The cost constraint problem in the general 
setting is trivial as in the case of 
classical information spectrum methods \cite{Han_book}.  
Namely, given a sequence $\chvec =\{\chn\}$ of channels 
$\chn:\cXn\rightarrow\cS(\cHn)$
as well as a sequence  $\costvec = 
\{\costn\}$ of functions $\costn : \cXn\rightarrow\bR$, 
which are called {\em cost functions}, and a real number $\gamma$, 
the capacity under cost constraint is nothing but 
the capacity $C(\chvec\restrict{\costvec ,\gamma})$ of 
the sequence of channels 
$\chvec\restrict{\costvec ,\gamma} =\{\chn\restrict{\costn,\gamma}\}$, where 
$ \chn\restrict{\costn,\gamma}$ is the 
restriction 
$\chn\restrict{\costn,\gamma}\,: 
\cXn_{\costn,\gamma}\ni\xn\mapsto
\chnarg{\xn}$
of the original channel $\chn$ to 
\begin{equation}
\label{def:cXncost}
  \cXn_{\costn,\gamma} \defeq 
\{ \xn \in\cXn \,|\, \costn (\xn)  \leq n \gamma \,\}. 
\end{equation}
In addition, the strong converse property in this case is represented as 
$C(\chvec\restrict{\costvec ,\gamma}) = 
C^\dagger (\chvec\restrict{\costvec ,\gamma})$. 
Needless to say, we can apply the general formulas in Theorem~\ref{thm:main} 
to these quantities. 

Now let us consider the situation where $\chvec =\{\chn\}$ is the 
stationary memoryless extension (\ref{cond:memoryless}) 
of $\ch :\cX\rightarrow\cS(\cH)$ 
and $\costvec =\{\costn\}$ is the 
additive extension 
\[ \costn (\xn) = \sum_{i=1}^n \cost (x_i), 
\] 
where $\cost$ is a function $\cX\rightarrow\bR$.  
We shall prove the following theorem, which was 
essentially obtained by Holevo 
\cite{Hol:quant-ph9809023, Hol:quant-ph9705054} except 
for the strong converse part. 

\begin{theorem}
In the stationary memoryless case with the additive cost, we have
\begin{equation}
\Label{cost_C=C+}
C(\chvec\restrict{\costvec ,\gamma}) 
=
\sup_{P\in\cP_{\cost,\gamma}(\cX)} I(P,\ch) ,
\end{equation}
where 
\[\cP_{\cost,\gamma} (\cX) 
\defeq 
\{ P\in\cP (\cX) \,|\, 
E_P[\cost] \defeq \sum_{x\in\cX} P(x) \cost (x) \leq \gamma \}. 
\]
If, in addition, $\dim\cH<\infty$ then the strong converse 
holds: $C^\dagger (\chvec\restrict{\costvec ,\gamma}) 
=
C(\chvec\restrict{\costvec ,\gamma})$. 
\end{theorem}

We first show that the (weak) converse part 
\begin{equation}
C(\chvec\restrict{\costvec ,\gamma}) \leq 
\sup_{P\in\cP_{\cost,\gamma}(\cX)} I(P,\ch) 
\label{weak_conv_memoryless_cost}
\end{equation}
is derived from the general formula. 
Let $\cPn\defeq\cP (\cXn_{\costn,\gamma})$ be 
the totality of probability distributions on 
$\cXn=\cXtn$ whose supports are finite subsets of 
\[ \cXnhat\defeq\cXn_{\costn, \gamma} =
\left\{ (x_1, \ldots ,x_n)\in\cXtn \,\left|\; 
\frac{1}{n}\sum_{i=1}^n c(x_i) \leq \gamma\,\right.\right\}.
\]
For any $\Pn\in\cPn$ and any permutation $\pi$ on 
$\{1,\ldots ,n\}$, $\Pn_{\pi}$ defined by $\Pn_{\pi}(x_1, \ldots , 
x_n)$ $=$ $\Pn(x_{\pi(1)},\ldots , x_{\pi(n)})$ also belongs to 
$\cPn$ and satisfies $I(\Pn, \chn)= I(\Pn_{\pi},\chn)$.  Since 
$I(\Pn, \chn)$ is concave with respect to $\Pn$, we can restrict 
ourselves to symmetric distributions when considering 
$\sup_{\Pn\in\cPn}I(\Pn,\chn)$.  
For a symmetric $\Pn\in\cPn$, the marginal distribution 
on $\cX$ belongs to $\cP_{\cost,\gamma}$ and satisfies
$I (\Pn, \chn) \leq n  I(P, \ch)$.  Hence we have 
\[
\sup_{\Pn\in\cPn} I (\Pn, \chn) 
\leq 
n \sup_{P\in\cP_{\cost,\gamma}} I(P, \ch),
\]
and (\ref{weak_conv_memoryless_cost}) follows from 
Lemma~\ref{lemma:infI=<liminf} and 
(\ref{generalC.1}) as in the costless case. 

Next, let us consider the direct part 
\begin{align}
\Label{cost_direct}
C(\chvec\restrict{\costvec ,\gamma}) 
\geq 
\sup_{P\in\cP_{\cost,\gamma}(\cX)} I(P,\ch) .
\end{align}
We use 
a slight modification of Lemma~\ref{lemma:direct} as follows. 
Let $P$ be a probability distribution in $\cP_{\cost,\gamma} (\cX)$
and $a$ be a real number. 
Given an arbitrary encoder $\encn:\{1,\ldots ,N\}\rightarrow
\cXtn$, let the decoder $\decn = \{\decnarg{1}, \ldots , \decnarg{N}\}$ be 
defined by 
\[ \decnarg{i} = \left( \sum_{j=1}^N \pi_j\right)^{-\frac{1}{2}} \pi_i 
\left( \sum_{j=1}^N \pi_j\right)^{-\frac{1}{2}}, 
\] 
where $\pi_i\defeq \spec{\chnarg{\encnarg{i}} - e^{na} {\charg{P}}^{\otimes n} > 0}$. 
It then follows from Lemma~\ref{lemma:ineq_for_direct} 
for $c=1$ (e.g.)\ 
that 
the average error probability of the code $\coden
 = (N, \encn, \decn)$ 
is bounded by
\begin{align*}
\Pe [\coden] 
&\leq \frac{2}{N}\sum_{i=1}^N 
\Tr \left[ \chnarg{\encnarg{i}} (I-\pi_i) \right] 
+
\frac{4}{N} \sum_{i=1}^N \sum_{j\neq i} 
\Tr \left[ \chnarg{\encnarg{i}} \pi_j\right] .
\end{align*}
Now let $\Ptn $ be the $n$th i.i.d.\ extension 
of $P$ 
and $\Pnhat \in\cP (\cXnhat)$
 be defined by
\begin{equation}
 \Pnhat (\xn) = 
\Ptn (\xn) / \Kn \quad\mbox{for}\quad \xn \in \cXnhat
=\cXn_{\costn, \gamma}, 
\label{def_Pnhat}
\end{equation}
where $\Kn\defeq \Ptn (\cXnhat)$.  Note that due to 
the assumption $P\in\cP_{\cost, \gamma}(\cX)$ and to the 
central limiting theorem we have
\begin{align}
\Label{limK}
\lim_{n\rightarrow\infty} \Kn 
\geq 
\lim_{n\rightarrow\infty} \Ptn \left(\cXn_{\costn, E_P[\cost ]}\right)
= 
\frac{1}{2}.
\end{align}
Generating the encoder $\encn$  
randomly according to the distribution 
\[ \Pnrc(\encn) = \Pnhat(\encnarg{1})\cdots\Pnhat(\encnarg{N}), 
\]
we see that 
there exists a code $\coden$ for $\chn\restrict{\costn,\gamma}$ of 
size $N$ satisfying
\begin{align*}
\Pe [\coden] & \leq 
2 \sum_{\xn\in\cXnhat} \Pnhat (\xn) 
\Tr\left[ \chnarg{\xn} 
\spec{\chnarg{\xn} -e^{na} {\charg{P}}^{\otimes n}\leq 0}\right] \\
&+ 
4 N \sum_{\xn\in\cXnhat} \Pnhat (\xn) 
\Tr\Bigl[
\bigl( \sum_{{\xn}'\in\cXnhat} \Pnhat ({\xn}') 
\chnarg{{\xn}'} \bigr) \, 
\spec{\chnarg{\xn} - e^{na} {\charg{P}}^{\otimes n} >0} \Bigr] \\
&\leq 
\frac{2}{\Kn} \sum_{\xn\in\cXnhat} \Ptn (\xn) 
\Tr\left[ \chnarg{\xn} \spec{\chnarg{\xn} -e^{na} {\charg{P}}^{\otimes n}\leq 0}\right] \\
&+ 
4 N \sum_{\xn\in\cXnhat} \Pnhat (\xn) 
\Tr\Bigl[
\bigl( \frac{1}{\Kn} {\charg{P}}^{\otimes n} \bigr) \, 
\spec{\charg{\xn} - e^{na} {\charg{P}}^{\otimes n} >0} \Bigr] \\
&\leq 
\frac{2}{\Kn} \sum_{\xn\in\cXtn} \Ptn (\xn) 
\Tr\left[ \charg{\xn} \spec{\charg{\xn} -e^{na} {\charg{P}}^{\otimes n}\leq 0}\right] 
+ \frac{4}{\Kn} N e^{-na} .
\end{align*}
Thus, letting $\Pvec=\{\Ptn\}$ and recalling (\ref{limK}) we have
\begin{equation}
C(\chvec\restrict{\costvec,\gamma}) 
\geq \infI (\Ptvec, \chvec) = I(P, \ch) ,
\label{C>=infI_cost}
\end{equation}
where the last equality follows from (\ref{infI=I}).  We have thus 
proved (\ref{cost_direct}). 

\begin{remark}
\rm
For the sequence $\vec{\hat{\bm{P}}}=\{\Pnhat\}$ defined 
from a $P\in\cP_{\cost,\gamma}(\cX)$ by (\ref{def_Pnhat}), 
the general formula (\ref{generalC.1}) implies that
\begin{align*}
C(\chvec\restrict{\costvec,\gamma}) 
&\geq 
\infI (\vec{\hat{\bm{P}}}, \chvec) \\
&=
\sup\, \{a\, \,|\, 
\lim_{n\rightarrow\infty} \sum_{\xn\in\cXnhat} 
\Pnhat(\xn) \Tr \left[\chnarg{\xn} \spec{\chnarg{\xn} - e^{na}\chnarg{\Pnhat} \leq 0}
\right] = 0\}
\\
&\geq
\sup\, \{a\, \,|\, 
\lim_{n\rightarrow\infty} \sum_{\xn\in\cXn} 
\Ptn(\xn) \Tr \left[\chnarg{\xn} \spec{\chnarg{\xn} - e^{na}\chnarg{\Pnhat} \leq 0}
\right] = 0\}
\\
&=
\infJ (\Ptvec, \sigmavec, \chvec) 
\quad\mbox{for}\quad
\Pvec=\left\{\Pn\right\}\;\;\mbox{and}\;\;
\sigmavec = \left\{\chnarg{\Pnhat}\right\},
\end{align*}
where the second 
inequality follows from (\ref{limK}).  
So, if we could use $\infI (\Ptvec, \chvec) =
\min_{\sigmavec}\infJ (\Ptvec,\sigmavec,\chvec)$, which 
is merely a conjecture at present (see  
Remark~\ref{remark:J>=I}), the inequality in (\ref{C>=infI_cost}) 
could be derived from the general formula as in the classical 
case. 
\end{remark}

Let us proceed to the proof of the strong converse part 
\begin{align}
\Label{cost_converse}
C^\dagger (\chvec\restrict{\costvec ,\gamma}) 
\leq 
\sup_{P\in\cP_{\cost,\gamma}(\cX)} I(P,\ch) 
\end{align}
under the assumption that $\dim\cH<\infty$. 
We claim that for any $\xvec\in\cXvechat=\{\cXnhat\}$, where 
$\cXnhat=\cXn_{\costn , \gamma}$, and any $\sigma\in\cS(\cH)$, 
\begin{equation}
\Label{ineq:supD=<sup_J}
\supD(\chvecarg{\xvec}\,\|\,\sigmatvec )
\leq 
\sup_{P\in\cP_{c,\gamma}(\cX)} J(P,\sigma,\ch) ,
\end{equation}
where $\chvecarg{\xvec}=\{\chnarg{\xn}\}$ 
and $\sigmavec =\{\sigmatensor\}$. 
We only need to show this for $\sigma$ such that 
$\mbox{Im}\,\sigma\supset\mbox{Im}\,\charg{x}$ 
for $\forall x\in\mbox{supp} (P)$, $\forall P\in\cP_{\cost,\gamma} (\cX)$, 
since the RHS is $\infty$ otherwise.  
For any $\xn \in\cXnhat$ and 
any real numbers $a$ and $0\leq s\leq 1$,  it follows from 
(\ref{ineq:OgaNag}) that 
\begin{align}
\Tr \left[ \chnarg{\xn} 
\spec{ \chnarg{\xn}  - e^{na} \sigmatensor >0} \right] 
&\leq \exp\Bigl[ -n \Bigl(as - \frac{1}{n} \sum_{i=1}^n
 \log \Tr \left[ \charg{x_i}^{1+s} \sigma^{-s}\right] \Bigr)
\Bigr] 
\nonumber 
\\
&\leq \exp\left[ -n \left(as - \psi(s)\right)\right] , 
\Label{ineq_for_cost_strongconv}
\end{align}
where 
\[
\psi(s) \defeq 
\sup_{P\in\cP_{\cost ,\gamma}(\cX)} 
\,\sum_{x\in\cX}\, P(x) \log \Tr \left[ \charg{x}^{1+s} \sigma^{-s}\right] .
\]
Let
\[
\cP_{\cost ,\gamma, 2}(\cX)\defeq 
\{P\,|\,P\in\cP_{\cost ,\gamma}(\cX) \;\;\mbox{and}\;\;
\left|{\rm supp}(P)\right|\leq 2\}, 
\]
where $\left|{\rm supp}(P)\right|$ denotes the number of elements of 
the support of $P$. 
Then a similar argument to section IV of 
\cite{Fuj-Nag} is applied to prove that 
 $\cP_{\cost,\gamma}(\cX)$ is the convex hull of $\cP_{\cost ,\gamma, 2}(\cX)$; 
see Appendix II.  Hence we have
\begin{align*}
\psi(s) &= \sup_{P\in\cP_{\cost ,\gamma, 2}(\cX)} 
\,\sum_{x\in\cX} \, P(x) \log \Tr \left[ \charg{x}^{1+s} \sigma^{-s}\right] .
\\
&= 
\max_{\omega\in\bar{\Omega}} g(s, \omega),
\end{align*}
where $\bar{\Omega}$ is the compact subset of $[0,1]\times\cS(\cH)^2$ 
defined as the closure of 
\[
\Omega=\left\{(\lambda, \charg{x_1}, \charg{x_2})
\, |\, 
0\leq\lambda\leq 1,\; (x_1, x_2)\in\cX^2,\; 
\lambda c(x_1) + (1-\lambda) c(x_2) \leq \gamma
\right\} ,
\]
and 
\[ g(s, (\lambda, \rho_1, \rho_2)) \defeq 
\lambda \log\Tr[\rho_1^{1+s} \sigma^{-s}] 
+ (1-\lambda) \log\Tr[\rho_2^{1+s} \sigma^{-s}].
\]
A similar argument to the
derivation of  (\ref{equivalent:supD=<maxD}) is applied 
to (\ref{ineq_for_cost_strongconv}) so that we have
\begin{align*}
&\lim_{n\rightarrow\infty} \Tr \left[ \chnarg{\xn} 
\spec{ \chnarg{\xn}  - e^{na} \sigmatensor >0} \right]  
=0 
\\
&\mbox{for}\;\; 
\forall a > 
\max_{\omega\in\bar{\Omega}} \frac{\partial}{\partial s} g(\omega, 0)
=
\sup_{P\in\cP_{c,\gamma}(\cX)} J(P,\sigma,\ch) , 
\end{align*}
which proves the claim (\ref{ineq:supD=<sup_J}). 
Now the strong converse (\ref{cost_converse}) is 
derived as follows:
\begin{align*}
C^\dagger (\chvec\restrict{\costvec ,\gamma})
& = 
\max_{\Pvec\in\cPvec(\cXvechat)} \; \min_{\sigmavec\in\cSvec(\cHvec)} \; 
\supJ (\Pvec, \sigmavec, \chvec)
\\ &\leq
\min_{\sigmavec\in\cSvec(\cHvec)} \; 
\max_{\Pvec\in\cPvec(\cXvechat)} \; 
\supJ (\Pvec, \sigmavec, \chvec)
\\
&=
\min_{\sigmavec\in\cSvec(\cHvec)} 
\max_{\xvec\in\cXvechat} \;
\supD (\chvecarg{\xvec}\,\|\,\sigmavec)  
\\ &\leq 
\min_{\sigma\in\cS (\cH)}
\max_{\xvec\in\cXvechat} \;
\supD (\chvecarg{\xvec}\,\|\,\sigmatvec)  
\quad\mbox{with}\quad
\sigmavec =\{\sigmatensor\}
\\ &\leq 
\min_{\sigma\in\cS(\cH)}\, 
\sup_{P\in\cP_{\cost,\gamma}(\cX)}
J(P,\sigma,\ch) 
\\ &=
\sup_{P\in\cP_{\cost,\gamma}(\cX)}\,
\min_{\sigma\in\cS(\cH)}\, 
J(P,\sigma,\ch) 
= \sup_{P\in\cP_{\cost,\gamma}(\cX)}\, 
I(P,\ch) ,
\end{align*}
where we have invoked the fact that similar relations to 
(\ref{capacity_memoryless.2}) hold for the present situation. 

\section{Concluding remarks}
\label{s9}
We have obtained a general formula for capacity of 
classical-quantum channels together with a characterization 
of the strong converse property by extending the information-spectrum 
method to the quantum setting.  The general results have been applied 
to stationary memoryless case with or without cost-constraint 
on inputs, whereby new simple proofs 
have been given to the corresponding coding theorems.  
Among many open problems concerning the present work, we would 
recall here only the following two; one is 
the problem mentioned in Remark~\ref{remark:J>=I} and 
the other is how to analyze (if possible) asymptotics of the quantum 
information spectrum directly, not by way of the theory 
of quantum hypothesis testing.  These problems will be important 
toward further 
developement of the quantum information-spectrum method. 

\section*{Acknowledgment}

The authors are grateful to an anonymous referee and 
Prof.\ T.S.\ Han for 
useful comments on the history of Feinstein's lemma. 

\bigskip
\bigskip
\appendix
\noindent {\large\bf Appendix I \quad Proof of Lemma \ref{lemma:sup_inf}}
\medskip

Let us begin with the attainability of $c\defeq\sup_{\fvec}\, [\fvec]^-_x$. 
We assume 
$-\infty < c<\infty$ first.  Then for every natural number $k$ 
there exists $\vec{f}^{(k)}=\{f^{(k)}_n\}_{n=1}^\infty\in\cFvec$ 
such that $[\vec{f}^{(k)}]^-_x > c- \frac{1}{k}$.  This implies that 
\[ 
\limsup_{n\rightarrow\infty} f^{(k)}_n \left(c-\frac{1}{k} \right)  
\leq x, 
\]
and hence there exists $n_k$ such that 
for any $n\geq n_k$, 
\[ f^{(k)}_n \left(c-\frac{1}{k} \right) \leq x+\frac{1}{k} . \]
Let us choose $\{n_k\}$ to satisfy $n_k < n_{k+1}$ ($\forall k$). Then  
every $n$ uniquely determines a number $k$ such that 
$n_k \leq n  < n_{k+1}$, which we denote by $k = k_n$.  Letting $f^*_n
\defeq f^{(k_n)}_n\in\cFn$ and $\fvec^* \defeq \{f^*_n\}_{n=1}^\infty\in\cFvec$, 
we have 
\[ f^*_n \left(c-\frac{1}{k_n} \right) \leq x+\frac{1}{k_n}. \]
This implies  that
$\limsup_{n\rightarrow\infty} f^*_n(c-\varepsilon) \leq x$ for 
any $\varepsilon >0$, and therefore we have 
$[\fvec^*]^-_x = c = \sup_{\fvec}\, [\fvec]^-_x$.  Next, let us consider the case when 
$c=\infty$.  Then for every natural number $k$ 
there exists $\vec{f}^{(k)}=\{f^{(k)}_n\}_{n=1}^\infty\in\cFvec$ 
such that $[\vec{f}^{(k)}]^-_x > k$, which implies the existence of 
a number $n_k$ such that for any $n\geq n_k$ we have $f^{(k)}_n (k) \leq x+\frac{1}{k}$. 
Then a similar argument to the previous one is applicable to construction of a 
sequence $\fvec^*= \{f^*_n\}\in\cFvec$ satisfying 
$\limsup_{n\rightarrow\infty} f^*_n (k) \leq x$ for 
any $k$, and therefore we have $[\fvec^*]^-_x = \infty = \sup_{\fvec}\, [\fvec]^-_x$.  
The remaining case $c=-\infty$ is trivial, since this means that 
$[\fvec]^-_x =-\infty$ for all $\fvec\in\cFvec$.  

Let us proceed to the attainability of $c\defeq\sup_{\fvec}\, [\fvec]^+_x,$. 
Assume $-\infty < c < \infty$.  Then for every $k$ there exists 
$\fvec^{(k)} = \{f^{(k)}_n\}_{n=1}^\infty \in\cFvec$ 
such that $[\vec{f}^{(k)}]^+_x > c- \frac{1}{k}$.  This implies that 
\[ 
\liminf_{n\rightarrow\infty} f^{(k)}_n \left(c-\frac{1}{k} \right)  
< x, 
\]
and hence there exists a $\delta_k >0$ such that the set 
\[ A_k\defeq \left\{n\,\left|\, f^{(k)}_n \left(c-\frac{1}{k} \right)  
\leq x -\delta_k\right.\right\}
\]
has infinitely many elements.   Let $\{B_k\}_{k=1}^\infty$ be a 
family of subsets $B_k\subset A_k$ such that $|B_k| = 
\infty$ and $B_k\cap B_l=\phi$ for $\forall k\neq\forall l$, and 
let $\fvec^*=\{f^*_n\}_{n=1}^\infty$ be defined by
\[
f^*_n = \left\{
\begin{array}{ccl}
f^{(k)}_n&\mbox{if}&n\in B_k , \\
 \mbox{an arbitrary element of $\cFn$}&\mbox{if}&
n\not\in \bigcup_k B_k .
\end{array} \right. 
\]
Then for every $k$ the set $\{n\,|\, f^*_n\left(c-\frac{1}{k}\right)\leq
x-\delta_k\}$ includes $B_k$ as a subset and hence has infinitely 
many elements.  This leads to
$\liminf_{n\rightarrow\infty} f^*_n\left(c-\varepsilon\right) < x$
for any $\varepsilon >0$, and therefore we have
$[\fvec^*]^+_x = c = \sup_{\fvec}\, [\fvec]^+_x$.  
The case $c=\infty$ can be proved similarly, and 
the case $c=-\infty$ is trivial.

Letting ${\cal G}_n$ be the set of monotonically 
nondecreasing functions $g_n(a)\defeq - \fn (-a)$ 
for $\fn\in\cFn$, we have
\[ 
\inf_{\fvec\in\cFvec}\, [\fvec]^-_x 
= -\sup_{\vec{g}\in\vec{\cal G}}\, [\vec{g}]^+_{-x}
\quad\mbox{and}\quad
\inf_{\fvec\in\cFvec}\, [\fvec]^+_x 
= -\sup_{\vec{g}\in\vec{\cal G}}\, [\vec{g}]^-_{-x} .
\]
The attainability of the infimums thus follows from that of 
the supremums. 

\bigskip
\bigskip

\noindent {\large\bf Appendix II \quad Proof that $\cP_{\cost,\gamma}(\cX)$ is 
the convex hull of $\cP_{\cost,\gamma, 2}(\cX)$}
\medskip

Let $P$ be 
an arbitrary distribution in $\cP_{\cost,\gamma}(\cX)$, and let $\cR_P$ denote the 
subset of $\cP_{\cost,\gamma}(\cX)$ consisting of all distributions $P'$ 
satisfying $E_{P'}[\cost] = E_P[\cost]$ and ${\rm supp}(P') 
\subset {\rm supp}(P)$.  
Since $\cR_P$ is convex and compact,  the element $P$ of $\cR_P$ 
can be represented as a convex combination of extreme points of $\cR_P$.  
Hence it suffices to show that the support of every extreme point of $\cR_P$ 
has at most two elements.  Suppose that 
a $P'\in\cR_P$ is written as $P'=\sum_{i=1}^k \lambda_i \,\delta_{x_i}$, where 
$\{x_1, \ldots, x_k\}={\rm supp}(P')$  and 
$\lambda_i\defeq P'(x_i) >0$.  If $k\geq 3$, there exists 
a nonzero real vector 
$(\alpha_1, \ldots, \alpha_k)\in\bR^k$ such that 
$\sum_{i=1}^k \alpha_i =0$ and $\sum_{i=1}^k \alpha_i \cost(x_i)=0$. 
Then, for a sufficiently small $\varepsilon >0$, 
$P_1 = \sum_{i=1}^k (\lambda +\varepsilon \alpha_i) \,\delta_{x_i}$ and 
$P_2 = \sum_{i=1}^k (\lambda -\varepsilon \alpha_i) \,\delta_{x_i}$
become two distinct distributions in $\cR_P$ and 
satisfy $P' = \frac{1}{2} (P_1 +  P_2)$, which means that 
$P'$ is not extreme.  Therefore, if $P'$ is an extreme point 
then $k=\left|\,{\rm supp}(P')\right| \leq 2$.

\end{document}